\documentclass[twocolumn]{aastex63}

\DeclareUnicodeCharacter{2212}{-}
\revised{\today}
\shorttitle{Time-resolved Optical Polarization Monitoring of 2MASS J21392676+0220226}
\shortauthors{Manjavacas et al.}
\graphicspath{{./}{figures/}}

\begin{document}

\title{Time-resolved Optical Polarization Monitoring of the Most Variable Brown Dwarf}

\author[0000-0003-0192-6887]{Elena Manjavacas}
\affiliation{AURA for the European Space Agency (ESA), ESA Office, Space Telescope Science Institute, 3700 San Martin Drive, Baltimore, MD, 21218 USA}
\affiliation{Department of Physics and Astronomy, Johns Hopkins University, Baltimore, MD 21218, USA}

\author[0000-0003-2446-8882]{Paulo A. Miles-P\'aez}
\affiliation{European Southern Observatory, Karl-Schwarzschild-Straße 2, 85748 Garching, Germany}

\author[0000-0001-7356-6652]{Theodora Karalidi}
\affiliation{Department of Physics, University of Central Florida, 4000 Central Florida Blvd., Orlando, FL 32816, USA}

\author[0000-0003-0489-1528]{Johanna M. Vos}
\affiliation{Department of Astrophysics, American Museum of Natural History, Central Park West at 79th Street, New York, NY 10024, USA}

\author[0000-0001-9300-7057]{ Max L. Galloway}
\affiliation{Department of Physics, University of Central Florida, 4000 Central Florida Blvd., Orlando, FL 32816, USA}

\author[0000-0001-8627-0404]{Julien H. Girard}
\affiliation{Space Telescope Science Institute, 3700 San Martin Drive, Baltimore, MD, 21218, USA}

\correspondingauthor{Elena Manjavacas}
\email{emanjavacas@stsci.edu}



\begin{abstract}

Recent atmospheric models for brown dwarfs suggest that the existence of clouds in substellar objects is not needed to reproduce their spectra, nor their rotationally-induced photometric variability, believed to be due to the heterogeneous cloud coverage of brown dwarf atmospheres. Cloud-free atmospheric models also predict that their flux should not be polarized, as polarization is produced by the light-scattering of particles in the inhomogeneous cloud layers of brown dwarf atmospheres. To shed light on this dichotomy, we monitored the linear polarization and photometric variability of the most variable brown dwarf, 2MASS~J21392676+0220226. We used FORS2 at the UT1 telescope to monitor the object in the $z$-band for six hours, split on two consecutive nights, covering one-third of its rotation period. We obtained the Stokes parameters, and we derived its time-resolved linear polarization, for which we did not find significant linear polarization (P = 0.14$\pm$0.07\%). We modeled the linear polarimetric signal expected assuming a map with one or two spot-like features and two bands using a polarization-enabled radiative-transfer code. We obtained values compatible with the time-resolved polarimetry obtained for 2MASS~J21392676+0220226. The lack of significant polarization might be due to photometric variability produced mostly by banded structures or small-scale vortices, which cancel out the polarimetric signal from different regions of the dwarf's disk. Alternatively, the lack of clouds in 2MASS~J21392676+0220226 would also explain the lack of polarization. Further linear polarimetric monitoring of 2MASS~J21392676+0220226, during at least one full rotational period, would help to confirm or discard the existence of clouds in its atmosphere.

\end{abstract}




\section{Introduction} \label{intro}

Brown dwarfs are substellar objects that cannot sustain hydrogen burning through their lifetimes as stars do \citep{Burrows}. Thus, brown dwarfs begin cooling after they {form}, evolving in spectral types as they age, from the late-M, {through the L spectral type (effective temperature, $\mathrm{T_{eff}}$ = 2400--1100~K)}, to the T ($\mathrm{T_{eff}}$ = 1100--600~K), and then Y ($<$600~K). Interestingly, when brown dwarfs reach a $\mathrm{T_{eff}}$ 1000--1100 K, at the transition between the L and T spectral types, their colors turn bluer in comparison to hotter brown dwarfs, in a very short range of temperatures (e.g. \citealt{Dupuy_Liu2012}). Atmospheric models have not completely explained yet which physical processes occur in the atmospheres of brown dwarfs at the L/T transition that turns brown dwarfs' spectra bluer as they cool down to the T spectral types. The first atmospheric models for substellar objects attempted to reproduce both L and T brown dwarf atmospheres simultaneously. L dwarf spectra were well-reproduced with cloudy models, by adding clouds of silicates to the photospheric spectra \citep{Jones_Tsuji1997}, and T-dwarfs were reproduced with clear atmospheres (cloud-free models). The transition between the L and the T spectral types was reproduced by a combination of cloudy and cloud-free models \citep{Allard2001, Ackerman_Marley2001, Saumon_Marley2008}. Later on, time-resolved photometric measurements found that brown dwarfs at the L-T transition presented higher variability amplitudes \citep{Radigan2014} than other spectral types, due, potentially, to heterogeneous clouds in brown dwarf atmospheres. This picture would be compatible with the idea that clouds break at the L/T transition. Nevertheless, some brown dwarf spectral characteristics are still not satisfactorily reproduced by traditional cloudy atmospheric models, such as low-surface gravity brown dwarf spectra continuum in the near-infrared (e.g. \citealt{Bonnefoy2014a, Manjavacas2014}), and the 10~$\mu$m molecular absorption feature \citep{Cushing2006, Suarez_Metchev2022}. 

In this context, \cite{Tremblin2016, Tremblin2017}  proposed new cloudless atmospheric models that assume a temperature gradient reduction caused by fingering convention. These models provide much better fits to the low-surface gravity brown dwarf spectra than cloudy models, and they explain most of the photometric variability due to temperature differences in the surface of brown dwarfs.
Cloudless models do not predict the existence of polarization in brown dwarfs {\citep{Tremblin2016}}.  The scattering of light produced by clouds particles are expected to be the main driver of linear polarization in brown dwarf atmospheres, but if clouds are not present, we do not expect to measure any polarization signal {\citep{Sengupta2001, Sengupta2003}}. 

Different groups have investigated the linear polarization properties of low-mass stars and brown dwarfs \citep{Menard2002, Zapatero_Osorio2005, Goldman2009, Zapatero-Osorio2011, Miles_Paez2013, Miles_Paez2015, Miles-Paez2017_pol,Manjavacas2017, Miles-Paez2019, Millar-Blanchaer2020}  finding polarization values in general $<$1\% at optical and near-infrared wavelengths. Nonetheless, few of them are time-resolved measurements, which would determine if there is a correlation with the rotational period of the object, confirming that the polarimetric measurements are definitely linked to heterogeneous cloud structures in the atmospheres of brown dwarfs. Potential rotation-induced polarization variability has been observed for a couple of objects (e.g. \citealt{Miles_Paez2015, Miles-Paez2017_pol, Millar-Blanchaer2020}). However, the correlation between rotational period (photometric variability) and polarimetric degree was only tentative.

In this paper we investigate the existence, degree, and variability amplitude of the linear polarization in the most highly photometrically variable brown dwarf, 2MASS J21392676+0220226 ({hereafter 2M2139+0220}), with the aim of further understanding the dynamics of its atmosphere and test the presence of clouds. 2M2139+0220 is a T1.5 brown dwarf \citep{Burgasser2006} ($J_{mag}$ = 15.3), and a member of the Carina-Near group (200$\pm$50~Myr) based on its full six-dimensional kinematics, and parallax \citep{Zhang2021}. 2M2139+0220 is a borderline object between brown dwarfs and planets, with an estimated mass of ${14.6}_{-1.6}^{+3.2}$ $\mathrm{M_{Jup}}$ \citep{Zhang2021}. 2M2139+0220 was found to be highly variable in the $J$-band by \cite{Radigan2012}, with a variability amplitude of $\sim$26\%, and a period of P = 7.721$\pm$0.005~hr. \cite{Radigan2012} showed that the most likely explanation for such high variability is the existence of patchy clouds that appear and disappear from the surface of the object as it rotates. Using \textit{Hubble Space Telescope} near-infrared data from the \textit{Wide Field Camera 3} instrument, \cite{Apai2013} further suggested that cloud thickness variations could be responsible for the high variability amplitude observed on its spectra. Later on, \cite{Apai2017} monitored this target using the \textit{Spitzer} telescope in eight different visits, finding a peak-to-peak variability amplitude of $\sim$11\% in the [3.6] band, and a period of $\sim$8.2~hr. 2M2139+0220 is observed edge on (i = 90$^{\circ}$, \citealt{Vos2017}). The different epochs of the 2M2139+0220 light curve show different patterns in each visit, further supporting the idea of variability due to changing cloud structures (spots and bands), as observed in giant planets of the Solar System. 

This paper is structured as follows: in Section \ref{observations_data_reduction} we describe the observations and data reduction. In Section \ref{light_curves} we present the $z$-band light curves in each ordinary and extraordinary beam, and their analysis, and  the polarization time-resolved measurement of 2M2139+0220. In Section \ref{modeling_pol} we compare our results with the results of the modeling of 2M2139+0220 atmosphere with a polarization enables radiative transfer code. {In Section \ref{other_pol_measurements} we compile all the polarimetric measurements from the literature with variability amplitude measurements}. Finally, in Section \ref{summary_remarks} we summarize our conclusions and explain our final remarks.


\section{Observations and Data Reduction} \label{observations_data_reduction}

We conducted linear polarimetric imaging photometry of 2M2139+0220 by using the Z$_{\rm SPECIAL+43}$ filter and the FOcal Reducer and low-dispersion Spectrograph 2 (FORS2; \citealt{Appenzeller1998}), which is mounted on the Antu unit (UT1) of the Very Large Telescope (VLT) of the European Southern Observatory (ESO) in Cerro Paranal, Chile. FORS2 is by default equipped with a detector system that is optimized for the red with a very low level of fringes thanks to a mosaic of two 2048$\times$4096 MIT CCDs (with 15 $\mu$m pixels). The plate scale is 0.252\arcsec pixel for the standard readout mode (2 × 2 binning). The polarization optics consists of a Wollaston prism as beam-splitting analyzer and a half-wave retarder plate to measure linear polarization. A mask with alternating transparent and opaque parallel strips avoids the overlapping of the ordinary and extraordinary rays from the Wollaston, yielding half of the field of view, which
is duplicated as ordinary and extraordinary rays in strips of size 3.4\arcmin$\times$11\arcsec. The central wavelength and the passband of the Z$_{\rm SPECIAL+43}$ filter are 916 and 18~nm, respectively. Observations were carried out during two half-nights starting on 20th and 21st October 2021 in delegated visitor mode (P.I. E. Manjavacas, 108.21ZK.001). Weather conditions were clear and the seeing was $\le$0.8\arcsec during {both} nights. Polarimetric images were acquired using {four} positions of the phase retarder
plate (0$^\circ$, 22.5$^\circ$, 45$^\circ$, and 67.5$^\circ$), with an exposure time of 120 sec per angle. One polarimetric cycle includes the four angles. To avoid as many systematics as possible, 2M2139+0220 was acquired on the same spot of the FORS2 detector 1 in both nights.  { To characterize the instrument efficiency and the instrumental polarization, one polarized (BD-13 5073) and one unpolarized  (WD 2149+021) standard stars were observed with the same instrumental configurations. We measured $P=0.09\pm0.02\%$ for WD 2149+021 and $P=3.24\pm0.12\%$, $\theta=147^{\circ}\pm1^{\circ}$, which are in agreement with the values tabulated in \cite{Fossati2007}. Thus, no corrections due to instrumental polarization were performed in our data.}


We reduced the Z$_{\rm SPECIAL+43}$ filter polarimetric data in imaging mode as follows: all $z$-band raw science data and flat images were bias subtracted using a previously median-combined bias {image}. Then we normalized all flat fields using their respective median values, and we median-combined them to obtain a master flat-field. We created a bad-pixel mask to mask bad and {dead} pixels. {Finally, to reduce} all polarimetric $z$-band science data, we divided all science polarimetric images by the master flat.

\section{Data Analysis and results}\label{light_curves}

\subsection{Differential photometry}\label{differential_photometry}

We performed aperture photometry on the reduced polarimetric images in the four positions of the half-wave retarder using the Python package \textit{aperture\_photometry} from {\textit{astropy.photutils}}. We used a circular aperture of 5 pixels, and a ring with an inner radius of  6 and an external radius of 10 pixels to estimate the value of the sky, on 13 different stars for calibration and the science target in the ordinary and extraordinary beams in the four positions of the half-wave retarder. {We did aperture photometry using apertures between 1 FWHM and 3 FWHM obtaining similar results.}

To derive the target's light curve by means of differential photometry, we followed a similar approach to \cite{Radigan2014}. We corrected the flux of the target independently in the four different rotation angles. First, we normalized all light curves to their median value. For each calibration star, a calibration light curve was created by median combining the light curves of all the other calibration stars. Then, the raw light curve of each calibration star was divided by its corresponding calibration light curve to obtain a corrected light curve.  To correct the target's flux, we chose the least intrinsically variable calibration stars. We followed the best-selection criteria for the calibration stars used by \cite{Radigan2014}, for which they subtracted from the corrected flux of each calibration star, a shifted version of itself, and divided it by $\sqrt{2}$ ($\sigma_{s}=[f_{cal}-f_{cal\_shifted}]/\sqrt{2}$). \cite{Radigan2014} then  identified  poor-quality calibration stars as those where $\sigma_{s}>1.5\times \sigma_{target}$. In this case, to be even more conservative, we used as calibration stars only those with $\sigma_{s}<\sigma_{target}$.
These criteria left  5 stars as good calibrators in the first night, and 6 in the second night. {As we did in \cite{Manjavacas2021}, we estimated the uncertainty for all the points in the target's light curve as the mean of the $\sigma_{s}$ of the light curve of the target and each of the selected calibration stars.}

After the flux of the target was corrected in the four angles independently using the same calibration stars, the standard deviation of all the calibration light curves was reduced significantly (by a factor between 2.0 and 10.0), implying that the correction of spurious fluctuations in their light curves was successfully removed. We multiplied the corrected light curves by the median of the initial non-corrected light curves of 2M2139+0220 in each angle of the retarder plate to have meaningful counts for each light curve. The corrected, non-phase folded light curves {for both nights} are shown in the Appendix in Fig. \ref{ord_extra_LC_night1} and \ref{ord_extra_LC_night2}.

In addition,   we calculated the $\tau$ Kendall correlation between the corrected target's light curve and the corrected light curves of calibration stars, to check if spurious variability was introduced in the correction process. In both nights, we found a weak to moderate correlation between the target's light curve and the calibration stars' light curve in the ordinary and extraordinary beams, probably due to residual effects of the airmass or seeing variations. The individual Kendall $\tau$ correlation for the two nights between those light curves can be found in Tables \ref{corr_cal_star_ord_night1}, and \ref{corr_cal_star_ord_night2} of the Appendix.

We performed a Bayesian Information Criterion (BIC) test to test the significance of the observed fluctuations, as in \cite{Naud2017}, \cite{Vos2020} and \cite{Manjavacas2021}. The BIC test is defined by \cite{Schwarz1978} as:

\begin{equation}
    \mathrm{BIC}=-2~\mathrm{ln}~\mathcal{L}_\mathrm{max} + k ~\mathrm{ln}~ N
\end{equation}

where $\mathcal{L}_\mathrm{max}$ is the maximum likelihood achievable by the model, $k$ is the number of parameters in the model {of your choice} and $N$ is the number of data points used in the fit. We calculated $\Delta\mathrm{BIC} = \mathrm{BIC}_{flat} - \mathrm{BIC}_{\mathrm{sin}}$ to assess whether a variable sinusoidal or non-variable flat model is favored by the data. {We used a sinusoidal model since it represents the simplest variable light curves that are commonly observed for variable brown dwarfs (e.g. \citealt{Radigan2014, Metchev2015, Apai2013, Biller2018, Vos2020})}. { The BIC penalizes the sinusoidal model for having additional parameters compared with the flat model.} If $\Delta\mathrm{BIC}$ is positive, then the sinusoidal model is preferred over the flat one, and the opposite if $\Delta\mathrm{BIC}$ is negative {\citep{Kass_Raftery1995}}. We calculated $\Delta\mathrm{BIC}$ for the eight individual light curves (ordinary and extraordinary in the four angles) in both nights separately.  We obtained that for all light curves the sinusoidal model was preferred over the flat light curve model. The only two exceptions were the ordinary light curve at the 45$^\circ$ angle, and the extraordinary light curve at 22.5$^\circ$ angle obtained in the second night. For these two light curves the flat model was slightly preferred over the sinusoidal one ($\Delta\mathrm{BIC}$ = -2.9 for the 45$^\circ$ angle ordinary beam, and $\Delta\mathrm{BIC}$ = -3.9 for the 22.5$^\circ$ angle extraordinary beam). The best fits to a sinusoidal and flat model can be found in Fig. \ref{ord_extra_LC_night1}, and \ref{ord_extra_LC_night2} in the Appendix {with the $\Delta\mathrm{BIC}$ values obtained for each individual light curve}. {Finally, we combined the ordinary and extraordinary light curves for each night at each position of the retarder plate for each night independently, and we run the BIC test again. We measured $\Delta\mathrm{BIC}$ values between 11.77 and 155.76, finding best fits for all light curves to a sinusoidal model, and further supporting that these light curves show significant variability. These light curves and the $\Delta\mathrm{BIC}$ values can be found in Fig. \ref{comb_ord_extra_LC_night12} of the Appendix.}

{We also run the BIC test on the entire non-normalized, non phase-folded ordinary and extraordinary light curves from night 1 and night 2 for the four angles of the retarder {plate}. From this analysis, we found that for all light curves the sinusoidal model was highly preferred over the flat light curve model, with $\Delta\mathrm{BIC}$ between 76.5 and 275.07. The BIC test estimated periods for these light curves that range between 8.23~hr and 10.43~hr for the limited amount of phase coverage of 2M2139+0220 that we have (about 1/3 of the light curve). The non-phase folded ordinary and extraordinary light curves, and the best fitting sinusoidal models for each of them with the individual values of the calculated $\Delta\mathrm{BIC}$, with their estimated amplitudes and periods are shown in the Appendix, Figures \ref{BIC_ord_LC} and \ref{BIC_extraord_LC}}.

\begin{figure*}[h]
    \centering
    \includegraphics[width=0.95\textwidth]{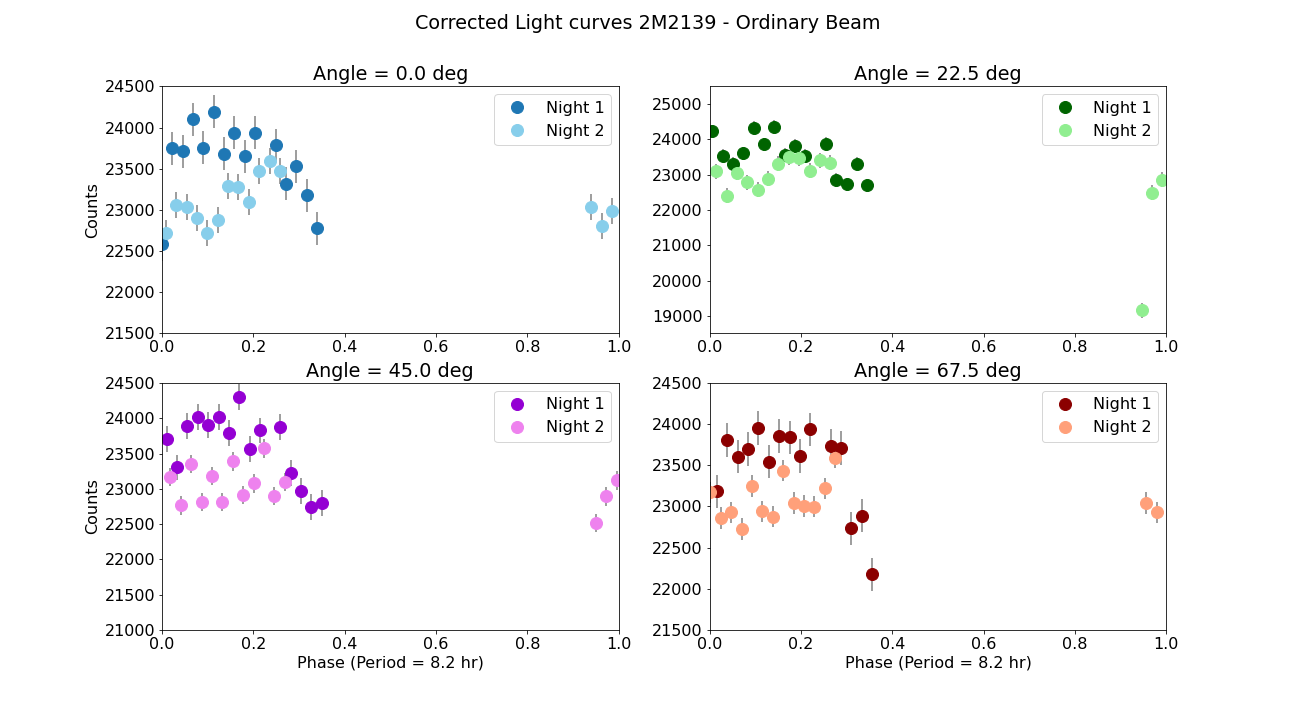}
    \includegraphics[width=0.95\textwidth]{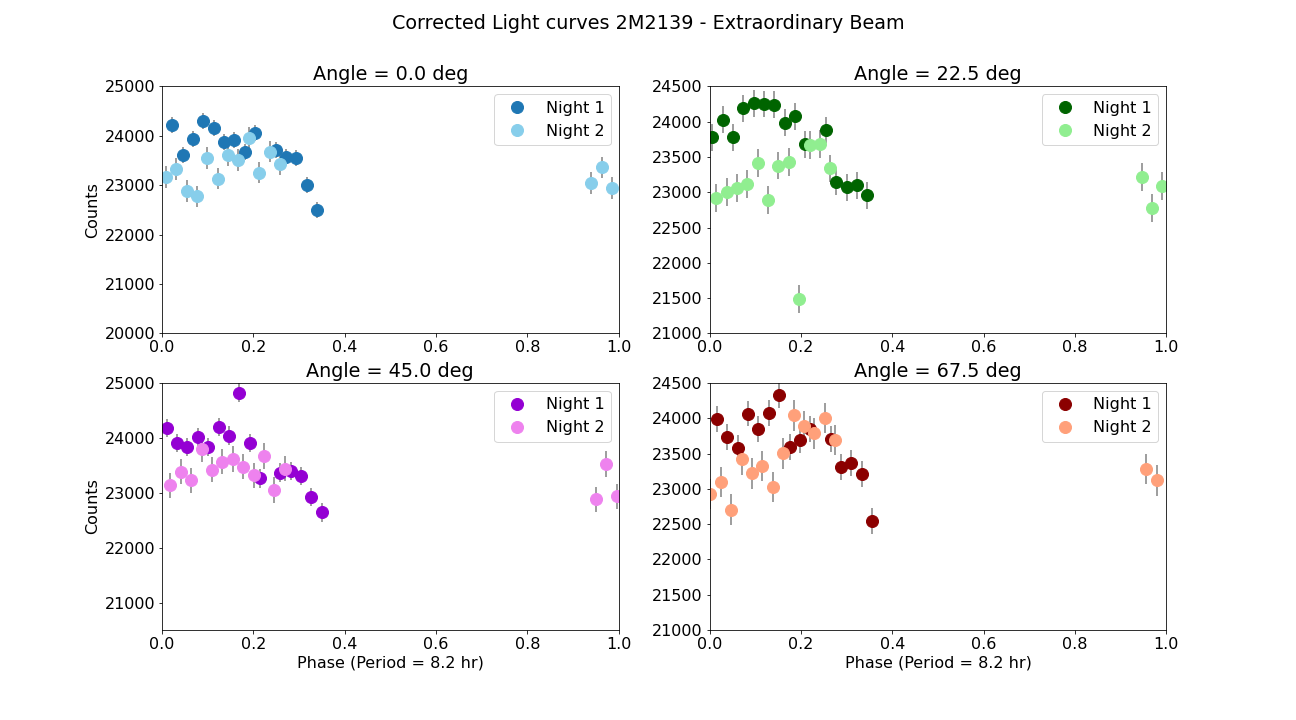}
   \caption{Phase folded light curves of 2M2139+0220 in the ordinary (top) and extraordinary beams (bottom) for angles 0.0, 22.5, 45.0 and 67.5$^\circ$ for nights 1 and 2. We show the face folded light curve using the latest 8.2~hr period from \cite{Apai2017}. {The absolute number of counts for the light curve in the two different nights are slightly different probably due to a difference in atmospheric conditions between the two nights.}}
    \label{ord_extra_LC_beams}
\end{figure*}

In Figure \ref{ord_extra_LC_beams}  we show the corrected and phase folded light curves (in counts) for 2M2139+0220  obtained in the ordinary and extraordinary beams in the four positions of the half-wave retarder plates. The absolute number of counts for the light curve on the two different nights is slightly different probably due to a difference in atmospheric conditions between the two nights. We phase-folded the eight light curves, and we run the BIC test again. To phase fold the light curves we used the rotational period reported in \cite{Apai2017} using \textit{Spitzer} light curves (P$\sim$8.2~hr) since they have the longest continuous time-coverage of all the light curves present in the literature ($\sim$28~hr of continuous light curve). {In addition, we found a better overlap between the light curves obtained in both nights using the 8.2~hr period than the 7.721$\pm$0.005~hr period reported in \cite{Radigan2012}, or the 7.83$\pm$0.1 hr reported by \cite{Apai2013}. 
The 8.2~hr period is closer to the  8.23--10.43~hr periods estimated by the BIC with the limited phase coverage we have for 2M2139+0220. In Fig. \ref{ord_extra_LC_beams_normalized}  of the Appendix, we show the normalized version of the phase-folded light curves using the 8.2~hr period. We also include a version of the normalized phase-folded light curve using the 7.721~hr rotational period in Fig. \ref{ord_extra_LC_beams_normalized_7.721} of the Appendix to show the improvement on the overlap of both light curves when using the 8.2~hr period over the 7.721~hr period}.

{We also phase-folded the ordinary and extraordinary combined light curve using the 8.2~hr rotational period (see Fig \ref{phase_folded_comb_ord_extra_LC_night12} of the Appendix). We run the BIC test for each independent light curve at each angle of the retarder plate, finding $\Delta\mathrm{BIC}$ values between 81.70 and 151.94, further supporting the presence of significant variability in the light curve.}

\subsection{Polarimetric analysis}\label{polarimetric_analysis}

\begin{figure*}[htp]
    \centering
    \includegraphics[width=0.98\textwidth]{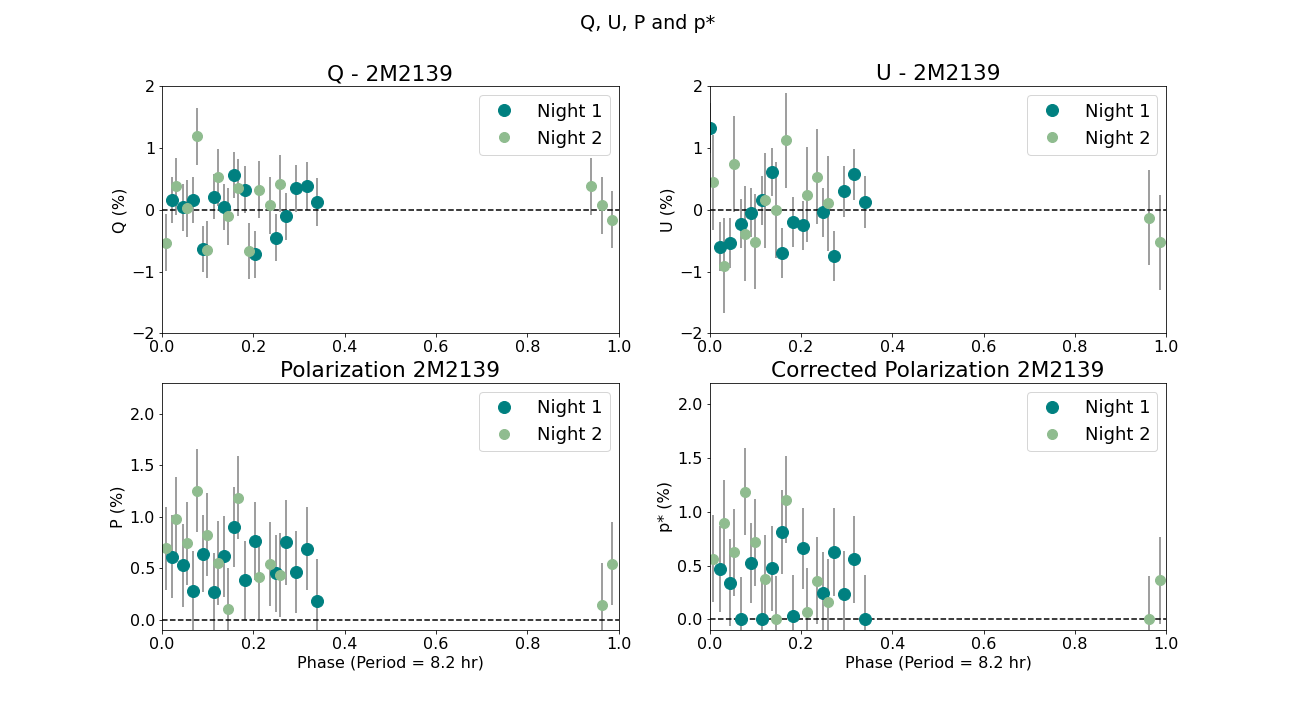}
   \caption{\textit{Q}, \textit{U}, polarization (\textit{P}), and corrected polarization (\textit{p*}) for nights 1 and 2 phase folded by the rotational period of the object. }
    \label{q_U_P_p_corr}
\end{figure*}

We calculated the Stokes parameters \textit{Q} and \textit{U}, and the linear polarization degree, \textit{P}, of our data by using the same circular aperture described in Section \ref{differential_photometry} and the flux-ratio method described in \cite{Zapatero_Osorio2005}. We obtained the Stokes parameters \textit{Q} and U, and the linear polarization degree,\textit{P}   following these equations:

\begin{equation}
    R_{q}^{2} = \frac{o(0.0)/e(0.0)}{o(45.0)/e(45.0)}
\end{equation}

\begin{equation}
    R_{u}^{2} = \frac{o(22.5)/e(22.5)}{o(67.5)/e(67.5)}
\end{equation}

\begin{equation}
    Q = \frac{R_{Q}-1}{R_{Q}+1}
\end{equation}

\begin{equation}
    U = \frac{R_{U}-1}{R_{U}+1}
\end{equation}

\begin{equation}
    P = \sqrt{Q^{2}+U^{2}}
\end{equation}

where \textit{o} is the {raw, uncorrected} flux of the ordinary beam, and \textit{e} refers to the {raw, uncorrected} flux of the extraordinary beam in the dual images of the single frames. Since linear polarization is always positive by definition, small values of \textit{P} and measurements affected by poor S/N are biased towards an overestimation of the true polarization. Thus, we correct the degree of linear polarization (eq. 7) by means of the following equation  \citep{wardle1974}:

\begin{equation}
    p* = \sqrt{P^{2}-\sigma_{P}^{2}}
\end{equation}

where $\sigma_{P}$ is the uncertainty on the linear polarization. In Figure \ref{q_U_P_p_corr} we show the measured Stokes parameters \textit{Q} and \textit{U}, the linear polarization, \textit{P}, and the debiased polarization, \textit{p*} using the ordinary and extraordinary light curves in both nights 1 and 2. {The uncertainties were formally propagated from the ordinary and extraordinary light curves in the four angles of the retarder plate}.

As performed earlier with the $z$-band photometric light curve of 2M2139+0220, we perform a Bayesian Information Criterion (BIC) to test if there are significant fluctuations in the measured corrected linear polarimetry, \textit{p*}. Calculating the BIC for \textit{p*}, we obtained a $\Delta$BIC~=~-1.69, indicating a {slight preference for the flat model of time-resolved corrected linear polarimetric light curve}, \textit{p*}. In conclusion,  although most of the ordinary and extraordinary light $z$-band light curves show significant variability, according to the BIC criteria, the corrected polarization \textit{p*} does not show significant variability, with the limited coverage that we have of 2M2139+0220's light curve (approximately one-third of the object's light curve).

\begin{figure}[h]
    \centering
    \includegraphics[width=0.5\textwidth]{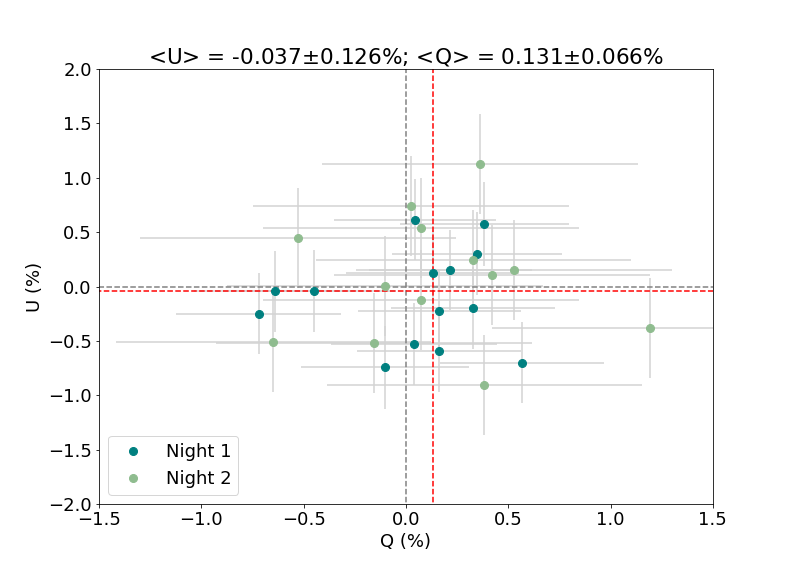}
   \caption{\textit{U}  vs \textit{Q} measured values. The dark green dots belong to the measurements of the first night, and the light green dots belong to the measurements of the second night. {The dashed red line shows the median \textit{Q} and \textit{U}, and the dashed grey line shows the \textit{Q} = 0\% and \textit{U} = 0\%.}}
    \label{U_vs_Q}
\end{figure}

Finally, we plot the \textit{Q} vs \textit{U}  values measured empirically in Fig. \ref{U_vs_Q} in both nights (dark green points for the first night data, and light gray for the second night data).  We measured a mean \textit{U}  value of -0.037$\pm$0.126\%, and a mean \textit{Q} value of 0.131$\pm$0.066\%. Using the mean values of \textit{Q} and \textit{U}, we obtained a mean polarization value of \textit{P}   = 0.14$\pm$0.07\%, and a corrected polarimetric value of \textit{p*} = 0.12$\pm$0.07\%, consistent with no significant linear polarization at the 3-$\sigma$ level in the phase covered by the light curve of 2M2139+0220.


\section{Modeling the expected polarization for 2M2139+0220}\label{modeling_pol}

Even though within the uncertainties of the observed \textit{P} we did not detect a significant polarization for 2M2139+0220, here we estimate the order of magnitude of the mean polarization, \textit{P}, predicted by a polarization enabled radiative transfer model \citep{2017A&A...607A..42S} with different surface map characteristics, and we compare it to the mean \textit{P} measured for 2M2139+0220 in Section \ref{polarimetric_analysis}. 

For this {purpose}, we used the polarization enabled radiative transfer code of \citet{2017A&A...607A..42S} to create a small grid of spectropolarimetric models of 2M2139+0220, and test if we can reproduce its observed mean linear polarimetric signal. The temperature, pressure, composition and cloud profiles of our model atmospheres were made using the radiative transfer code of \cite{Ackerman_Marley2001} and \cite{Marley_Sonora_bobcat2021}. Following the {derived} cloud properties of 2M2139+0220 from \citet{Apai2013}, our model atmospheres have two cloudy models: a background atmosphere with $T_\mathrm{eff}=1,100$K and optically thicker clouds ($f_\mathrm{sed}=1$), and a single model for all cloud features with $T_\mathrm{eff}=1400$~K and optically thinner clouds ($f_\mathrm{sed}=4$). The published maps of 2M2139+0220 include only spot-like features \citep{Apai2013,Karalidi2015}, but they were all made before long temporal baseline observations of L/T dwarfs showed that they can have banded cloud structures like Jupiter's belts and bands (\citealt{Apai2017, Millar-Blanchaer2020}). As we only observed one-third of the full rotational period of 2M2139+0220, we did not aim to accurately map the atmosphere of 2M2139+0220, but to test if any model can reproduce the observed {mean} polarization of the time-resolved FORS2 $z$-band linear polarimetry data. Therefore, we did not perform an exhaustive modeling effort, and we limited our models to include two bands and one {or two} cloud spots, as suggested by \cite{Apai2017} for 2M2139+0220 after mapping the object using more than 28~hr of Spitzer continuous photometric data (see \citealt{Apai2017} for further details). 

We tested different sizes of cloud spots ranging from small circular spots ($5^\circ$ radius) to oval spots comparable to Jupiter's Great Red Spot (GRS). {A detailed list of the models we ran is shown in the Appendix in Table~\ref{tab:themodelruns}.} Fig.~\ref{plot:pol_model} shows our modelled light curve for a model with a GRS-like spot, extending $30^\circ$ in longitude and $20^\circ$ in latitude, on the south hemisphere and two symmetric bands around the equator (solid, black line) against our time-resolved $z$-band linear polarimetry observations of 2M2139+0220. We also show a model with a circular equatorial spot with a diameter of $10^\circ$ (dashed, blue line). {Finally, we show a model with two spots: a GRS-like on the northern hemisphere and the $10^\circ$ spot on the southern hemisphere, with 79$^\circ$ in longitude separating the two (dashed dotted, red line). To understand the effect of the different features on the phase curve, Fig.~\ref{plot:pol_model_extended} represents our modeled \textit{P} for the two-spot model of Fig.~\ref{plot:pol_model}, and the corresponding maps for a full rotational period. The banded structures cause a constant \textit{P} as they are rotationally symmetric. The spots on the other hand, are rotationally non-symmetric and are responsible for all of the observed phase variability.} 

The single GRS-like spot or the two-spot model match better the range of the observed mean \textit{Q} and \textit{P} of 2M2139+0220, while the smaller spot shows a larger, near-constant \textit{Q} and \textit{P} at all phase angles. Both single-spot models show low phase fluctuations in \textit{U}, while the two-spot model shows larger phase fluctuations in \textit{U}. Finally, the two-spot model also shows larger phase fluctuations in \textit{Q} and \textit{P}. {While the uncertainties of our observations are too large for us to prefer a given model, the mean observed \textit{P} shows relatively large phase fluctuations that might be better matched by the two than the one-spot models. These large phase fluctuations might potentially hint at a more complex map for 2M2139+0220, in agreement with flux-only observations. 
In any case, to be able to decide which model reproduces best the data, further linear polarimetric monitoring covering at least one full rotational period of 2M2139+0220 is needed.} 

{Finally, we note that a direct comparison of the model unpolarized flux against archival light curves is not possible as observations suggest that the light curve of 2M2139+0220, like all L/T atmospheres, evolves significantly over time \citep{Radigan2012,Apai2013}. Additionally, archival observations are all in the $J-$band and our observations are in the $z-$band. These bands probe different pressures ($\delta$P$\sim$8bar) in the atmosphere of 2M2139+0220. Observations of other targets show that different bands exhibit different light curves (\citealp{Biller2013}, \citealp{Yang2016}, or compare the light curve shapes of \citealp[][]{Buenzli2015} and \citealp{Apai2021}) due to different cloud structures in the different atmospheric layers of brown dwarfs. Tentatively though, we can use the amplitude of the model flux light curves to compare against the amplitude of past observations and test whether some models are preferred over others. In agreement with the \textit{Q} and \textit{P} models, the smaller single-spot models can be tentatively excluded due to the very small light curve amplitude they produce ($\ll1$\% in the $z-$ and $\lesssim3$\% in the $J-$band). On the other hand, a single GRS-like spot (amplitude of $\sim12$\%) and the two-spot models (amplitudes of $12$\% to $25$\% in the $z-$ and $J-$bands) are all plausible for 2M2139+0220. }


\begin{figure*}[h]
    \centering
    \includegraphics[trim={1cm 0cm 0cm 0cm}, width=0.7\textwidth]{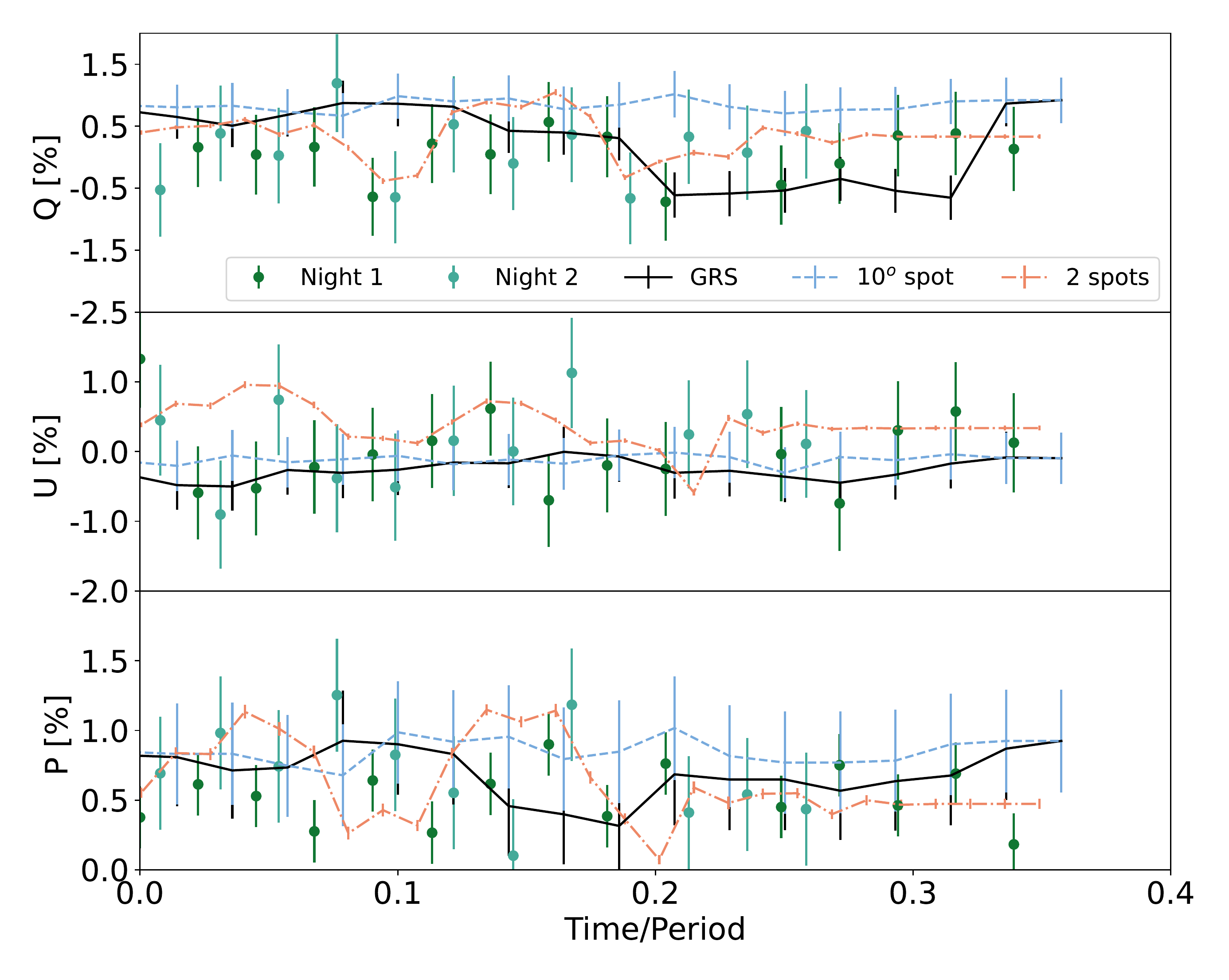}
   \caption{Observed \textit{Q} (top panel), \textit{U} (middle panel) and \textit{P}   (bottom panel) against model light curves of atmospheres with: two bands and a GRS-like spot ($30^\circ$ in longitude by $20^\circ$ in latitude) extension) on the southern hemisphere (solid, black line), an equatorial circular spot with $10^\circ$ diameter (dashed, light blue line), or two spots one on the north and one on the south hemisphere (dashed-dotted, orange line). }
    \label{plot:pol_model}
\end{figure*}

\begin{figure*}
    \centering
    \includegraphics[width=0.95\textwidth, trim={.6cm 0cm 0cm 0cm}]{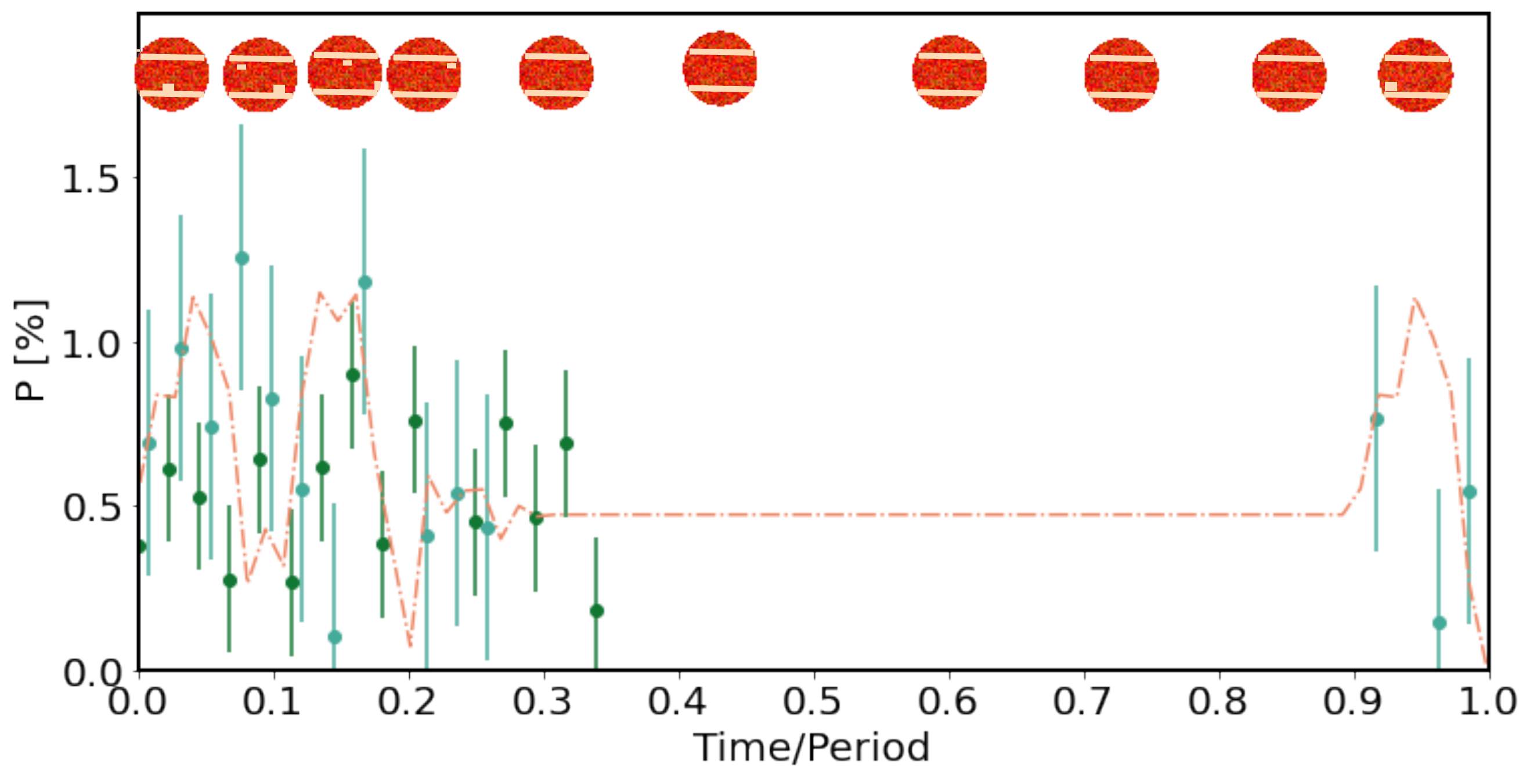}
   \caption{An extended version of Fig.~\ref{plot:pol_model} showing the full model light curve of the two spot model against our observations. The corresponding top-of-the-atmosphere maps are shown at the top. Bands in our model atmosphere result in a non-zero constant polarization (see phase angles of 0.4 to 0.9). The cloud spots on the other hand are responsible for the phase variability of our model \textit{P}. } 
    \label{plot:pol_model_extended}
\end{figure*}

\section{Polarization measurements for other variable brown dwarfs}\label{other_pol_measurements}

\begin{figure*}[h]
    \centering
    \includegraphics[width=0.85\textwidth]{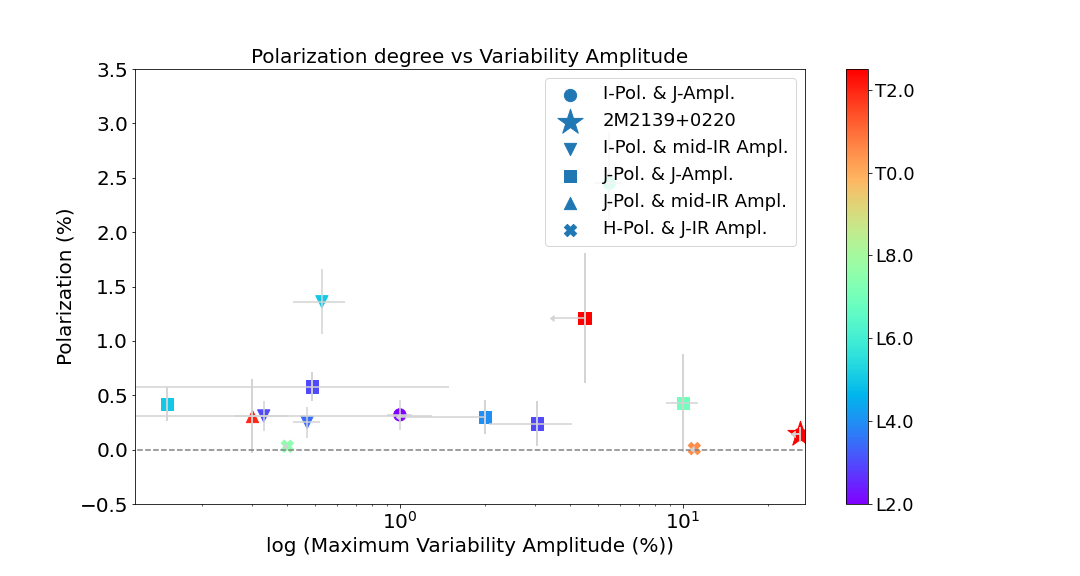}
   \caption{Maximum variability amplitudes measured for a sample of 15 brown dwarfs vs their linear polarization values. For the majority of these objects both variability amplitudes and linear polarization were measured separately at different times and by different instruments. The colors denote spectral types as indicated in the color bar in the left hand side of the plot.}
    \label{plot:pol_vs_var}
\end{figure*}

We collected the maximum values of polarization in the $I$-, $J$-, and {$H$-bands}, and the maximum values of photometric variability in the $J$-band, and in the mid-infrared from the literature for those brown dwarfs for which photometric variability and linear polarization were measured at some point. We collected data for 15 brown dwarfs in total, with spectral types between L2.0 and T2.5. The polarization values and the variability amplitudes are in most of the cases not measured simultaneously, and in general they were measured by different instruments, and published by different groups. Thus, although these measurements might provide some hints on how variability amplitude and linear polarization might be correlated for this reduced sample of objects, ideally, we will need \textit{simultaneous} time-resolved measurements of photometric variability and linear polarization \textit{in the same bands} to be able to compare these measurements with the measurements of 2M2139+0220 presented in this work, and to derive more definitive conclusions for the relation between linear polarimetry and photometric variability. In Fig. \ref{plot:pol_vs_var} we show only the maximum variability amplitude registered in the literature for each object (in the near-infrared {or} in the mid-infrared), versus its maximum polarization measured (in the $I$-band, $J$-band, {$H$-band} or $z$-band for 2M2139+0220). Since not all studies showed the corrected polarization, \textit{p*} for the objects, we only show the linear polarization, \textit{P}. Some objects have various measurements of variability amplitudes or polarization signals, but we just register the maximum of both. In Table \ref{table:pol_vs_var} and Fig. \ref{plot:pol_vs_var} we show the brown dwarfs with variability amplitudes and linear polarization measured in the literature. In Fig. \ref{plot:pol_vs_var} we color coded the measurements by spectral types.

As observed in Fig.  \ref{plot:pol_vs_var}, there is not obvious trend between photometric variability amplitude and the degree of linear polarimetry, \textit{P}, with the limited sample for which we could find measurements of photometric variability and linear polarization in the literature.  Nevertheless, to confirm this hypothesis, we would need to monitor \textit{simultaneously} for variability and for linear polarization all the targets in Fig. \ref{plot:pol_vs_var} during at least one rotational period in the same band. If we find a general correlation for most of the objects between the variability in linear polarization and the photometric variability, then the existence of heterogeneous clouds in the atmospheres of brown dwarfs producing both linear polarization and photometric variability might be the most likely explanation. 

An alternative possibility is that we observe photometric variability due to heterogeneous cloud structures, but no changes in linear polarization, as for 2M2139+0220 presented here, and as suggested by Fig. \ref{plot:pol_vs_var}. In this case, the photometric variability might be {potentially} produced by banded structures which are rotationally symmetric across the disk of the object \citep[e.g., Luhman~16][and Fig.~\ref{plot:pol_model_extended}]{Millar-Blanchaer2020} or by many smaller scale vortices \citep{2021ApJ...923..113M}. Thus, when the disk-integrated linear polarization is measured, the polarization values are low {and do not show phase variability}. {Finally, the fast rotation of brown dwarfs is expected to cause non-zero oblateness for some of these atmospheres which would also result in a small, non-zero polarization even for a cloud-decked (without cloud heterogeneities) atmosphere \citep{marley2011,Miles_Paez2013, chakrabarty2022}. As the oblateness of the atmosphere does not change in a rotational period, the lack of cloud heterogeneities would result in a constant polarization across all phase angles. We note, however, that the rotational period of 2M2139+0220 is large and the oblateness of this atmosphere alone is expected to cause a polarization of $<$ 0.01\% (see \citealp{marley2011} their Fig. 9 or the predicted polarization of Luhman 16A which has a comparable rotation period to 2M2139+0220 in \citealp{Millar-Blanchaer2020}).}

Another possibility is, as suggested by \cite{Tremblin2015, Tremblin2016, Tremblin2017} the lack of clouds in most brown dwarf atmospheres. In this case, we would not expect to measure any linear polarization for most brown dwarfs. To be able to probe any of these scenarios, further time-resolved linear polarimetric measurements with simultaneous time-resolved photometry for a more extended sample of brown dwarfs are needed.

\begin{table*}[h]
	\caption{Brown dwarfs in the literature with measurements of linear polarization and photometric variability in different bands. We include only the maximum polarizations signal and variability amplitudes in different filter.}  
	\label{table:pol_vs_var}
	\centering
	\small
	\begin{center}
		\begin{tabular}{llllll}
        \hline
		\hline 
			
Name & SpT  &\textit{P}   (\%)   & A (\%) $J$-band & A (\%) mid-IR & References \\  
\hline
2MASS J00452143+1634446 & L2.0 & 0.32$\pm$0.14 ($I$-band)  & 1.00$\pm$0.10 & & (1) (5)	\\
2MASS J03552337+1133437 &L5.0   & 0.42$\pm$0.16 ($J$-band)  & $<$0.15 & & (2) (priv. comm)\\
2MASS J05012406--0010452 & L4.0  & 0.30$\pm$0.16 ($J$-band) & 2.00$\pm$1.00 & & (2) (6) \\
2MASS J17260007+1538190 & L3.0  & 0.58$\pm$0.13 ($J$-band) &  &$0.49$ & (2) (7) \\
PSO J318.5338-22.8603   & L7.0  & 0.43$\pm$0.45 ($J$-band) & 7.00$\pm$1.00 & & (2) (8)  \\
2MASS J22081363+2921215 & L3.0  & 0.24$\pm$0.21 ($J$-band) &$<$3.00 & & (2) (9) \\
2MASS J00361617+1821104 & L3.5 & 0.25$\pm$0.14  ($I$-band) &  & 0.47$\pm$0.05 & (1) (7) \\
2MASS J15074769--1627386& L5.0 & 1.36$\pm$0.30 ($I$-band)   & & 0.53$\pm$0.11 & (1) (7)\\
2MASS J17210390+3344160 & L3.0 & 0.31$\pm$0.14 ($I$-band)   & & 0.33$\pm$0.07 & (1) (7)\\
2MASS J22443167+2043433 & L6.5 & 2.45$\pm$0.47 ($I$-band)  & 5.50$\pm$0.60 & & (1) (6)\\
2MASS J12545393-0122474 & T2.0 & 0.31$\pm$0.34 ($J$-band) &	&$<$0.30 & (3) (7) \\
2MASS J01365662+0933473 & T2.5  & 1.21$\pm$0.60 ($J$-band) & $<$4.50 & & (4) (10)\\
2MASS J2139+0220 & T2.5 & 0.14$\pm$0.07 ($z$-band) &   $<$26 & & (This work) (11)\\
{Luhman 16A} & L7.5 &  0.031$\pm$0.004 ($H$-band) & $<$0.4 & & (12) (13)\\
{Luhman 16B} & T0.5 &  0.010$\pm$0.004 ($H$-band) & $\sim$11 & & (12) (13) \\
			\hline			
		\end{tabular}
	\end{center}
	\footnotesize{References: (1) \cite{Zapatero_Osorio2005}; (2) \cite{Miles-Paez2017_pol}; (3) \cite{Miles_Paez2013}; (4) \cite{Zapatero-Osorio2011}; (5) \cite{Vos2020}; (6) \cite{Vos2018}; (7) \cite{Metchev2015}; (8) \cite{Biller2015}]; (9) \cite{Manjavacas2021}; (10) \cite{Apai2013}; (11) \cite{Radigan2012}; (12) \cite{Buenzli2015}; (13) \cite{Millar-Blanchaer2020}}\\
\end{table*}

\section{Summary and Final Remarks}\label{summary_remarks}

{We monitored 2M2139+0220 for linear polarimetry in the $z$-band during $\sim$3~hr continuously in two different consecutive nights. 
We obtained the light curve of 2M2139+0220 in the ordinary and extraordinary beams at the four polarization angles of the Wollaston prism (8 light curves in total, 16 in total for the two nights). 
After correcting these light curves using non-intrinsically variable calibration stars on the field of view, we found that the eight light curves showed significant variability using a BIC analysis (see Section \ref{differential_photometry}). Of the 16 individual light curves obtained during the two nights, 14 showed significant variability. The periods estimated by the BIC analysis with the limited amount of phase coverage we achieved, expand between 8.23~hr and 10.19~hr, which is marginally compatible with the $\sim$8.2~hr period measured by \citep{Apai2017}.}

{Using the ordinary and extraordinary light curves at the four angles of the retarder plate, we measured \textit{Q}, \textit{U}, \textit{P} and the corrected polarization \textit{p*} following equations (2) to (7).  We obtained the median value of \textit{Q} = 0.131$\pm$0.066\% and \textit{U} = -0.037$\pm$0.126\% in the limited phase coverage we had for the 2M2139+0220 light curve. We obtained a mean polarization value of \textit{P}   = 0.14$\pm$0.07\%, and a corrected polarimetric value of \textit{p*} = 0.12$\pm$0.07\%, consistent with no significant linear polarization at the 3-$\sigma$ level in the phase covered by the light curve of 2M2139+0220.} {Comparing these values of \textit{Q}, \textit{U}, and \textit{P} with the only other brown dwarf with similar time-resolved linear polarimetric data,  the much brighter Luhman~16A, we see that it also shows low level of polarimetric signal  (\textit{P} = 0.031$\pm$0.004\%) in the near-infrared, a \textit{Q} = −0.007\%, and  \textit{U} = 0.027\%, due most likely to banded structures \citep{Millar-Blanchaer2020}. }

In Fig. \ref{plot:pol_vs_var} we compiled all brown dwarfs with spectral types between L2.0 and T2.5 which have measurements of variability amplitude in one or several bands, and measurements of linear polarization. Although most of these measurements of maximum variability amplitudes, and maximum linear polarization were not obtained simultaneously as for 2M2139+0220,  and not necessarily in the same bands,  for the limited sample that we contemplate here, a priori, there is no obvious correlation between high variability amplitudes and high linear polarization signal. 
Nevertheless, to confirm this hypothesis, we would need to monitor \textit{simultaneously} for variability and for linear polarization all the targets in Fig. \ref{plot:pol_vs_var} during at least one rotational period in the same band.

To test whether clouds could reproduce the {order of magnitude of the mean polarization we attempted to measure, we created a number of model atmospheres and calculated their polarization in the $z$-band}. As our observations were phase-limited we could not use the observations to accurately map 2M2139+0220. In particular, the limited phase coverage of our observations does not allow us to break the degeneracy of banded structures with cloud spots for which we need longer phase coverage \citep[e.g.,][]{Apai2017}. Thus, based on models of past flux-only observations, we modeled atmospheres with two bands and one or two spots. {The \textit{Q} and \textit{P} of our model atmospheres containing a single large spot, or two spots {are consistent with} the mean observed \textit{Q} and \textit{P}. This suggests that cloud structures in the atmosphere of 2M2139+0220 could potentially be responsible for the mean polarization measured, even though we do not measure significant polarization at the 3-$\sigma$ level.}
{Additionally, the observed mean \textit{P} might potentially hint at a more complex map for the atmosphere of  2M2139+0220, in agreement with past mapping efforts of L/T brown dwarfs that showed that these atmospheres have complex maps \citep{Apai2013, Crossfield2014, Apai2017}. }

{While the uncertainties of our observations are too large for us to find a best-fit model, Figs.~\ref{plot:pol_model} and~\ref{plot:pol_model_extended} show that the \textit{Q}, \textit{U} and \textit{P} phase curves hold information about the cloud map of our model atmospheres. Future observations that can reduce the observational uncertainty to the 0.01\% level or lower (versus {$\sim$0.4\%, per bin here}) would allow us to separate these models and also exclude the possibility that oblateness is responsible for the observed mean polarization. Such an observation would allow us to use our models to map the atmosphere of 2M2139+0220 and break the degeneracies that flux-only maps have by distinguishing spots on the north versus the south hemisphere of the atmosphere. }

2M2139+0220 is probably the best-case scenario to test the existence of linear polarization due to scattering by the cloud particles in its atmosphere since it is a relatively bright, highly variable brown dwarf, that is observed edge-on \citep{Vos2017}. Thus,  the cloud structures and heterogeneities producing high linear polarization are likely observable, and they would likely produce an observable polarimetric signal correlated with its light curve. The linear polarization signal expected for other brown dwarfs with less favorable configurations might be much weaker depending on their configuration, which might explain why high polarimetric signals ($>$1\%) {have been reported for few brown dwarfs}  \citep{Zapatero_Osorio2005, Zapatero-Osorio2011, Miles_Paez2013, Miles_Paez2015, Manjavacas2017}, and why the measured linear polarimetry for the same brown dwarf is different from epoch to epoch \citep{Miles-Paez2017_pol}. {In addition, the polarization signal might change across one or several rotational periods, since the atmospheres of brown dwarfs evolve with time \citep{Apai2017}. Thus, to properly characterize the linear polarization signal, and its evolution, at least one full rotational period needs to be monitored \citep{Miles-Paez2017_pol}}.

To confirm the expected variability for the linear polarimetric signal of 2M2139+0220, further polarimetric continuous monitoring during at least one full rotational period is needed. In addition, {time-resolved} polarimetric monitoring of other highly variable brown dwarfs, like VHS~1256--1257 b, PSO 318.5--22, or 2MASS~J00470038+6803543, would help to further confirm the existence of clouds in the atmospheres of brown dwarfs, and learn about the properties of the particles in their atmospheres.

\acknowledgments

We thank the anonymous referee for the useful comments provided which helped to improve our manuscript. 
Based on observations collected at the European Organisation for Astronomical Research in the Southern Hemisphere under the ESO programme 108.21ZK.001. TK and MLG acknowledge the support of NASA Exoplanets Research Program grant 80NSSC21K0396.\\

\facilities{European Southern Observatory (ESO)}


\software{astropy \citep{2013A&A...558A..33A}}



\bibliography{Polarization_2M2139}{}
\bibliographystyle{aasjournal}

\newpage

\appendix


\begin{table*}[h]
	\caption{{Description of the models used in this paper. Our models had one or two spots located at different latitudes in the atmosphere and either had no bands or had a North and a South band with 22$^\circ$ latitudinal extent. The central latitude of each spot and band is also given.}}  
	\label{tab:themodelruns}
	\centering
	\small
	\begin{center}
		\begin{tabular}{lllll}
        \hline
		\hline 
			
Number of spots & spot size  &  Spot location ([N]orth  & Bands  & Band location  \\  
 &  [deg]  &  or [S]outh) [deg]  & [Y/N] & [deg] \\ 
\hline
1  &  10$\times$10 & 0 & N & N/A \\
1  &  20$\times$10 & 0 & N & N/A \\
1  &  20$\times$20 & 0 & N & N/A \\
1  &  30$\times$10 & 0 & N & N/A \\
1  &  30$\times$20 & 0 & N & N/A \\
1  &  10$\times$10 & 28 [N] & N & N/A \\
1  &  20$\times$10 & 28 [N] & N & N/A \\
1  &  20$\times$20 & 28 [N]& N & N/A \\
1  &  30$\times$10 & 28 [N]& N & N/A \\
1  &  30$\times$20 & 28 [N]& N & N/A \\
1  &  10$\times$10 & 28 [S] & N & N/A \\
1  &  20$\times$10 & 28 [S] & N & N/A \\
1  &  20$\times$20 & 28 [S]& N & N/A \\
1  &  30$\times$10 & 28 [S]& N & N/A \\
1  &  30$\times$20 & 28 [S]& N & N/A \\

1  &  10$\times$10 & 0 & Y & 44 [N \& S]\\
1  &  20$\times$10 & 0 & Y & 44 [N \& S]\\
1  &  20$\times$20 & 0 & Y & 44 [N \& S]\\
1  &  30$\times$10 & 0 & Y & 44 [N \& S]\\
1  &  30$\times$20 & 0 & Y & 44 [N \& S]\\
1  &  10$\times$10 & 28 [N] & Y & 44 [N \& S]\\
1  &  20$\times$10 & 28 [N] & Y & 44 [N \& S]\\
1  &  20$\times$20 & 28 [N]& Y & 44 [N \& S]\\
1  &  30$\times$10 & 28 [N]& Y & 44 [N \& S]\\
1  &  30$\times$20 & 28 [N]& Y & 44 [N \& S]\\
1  &  10$\times$10 & 28 [S] & Y & 44 [N \& S]\\
1  &  20$\times$10 & 28 [S] & Y & 44 [N \& S]\\
1  &  20$\times$20 & 28 [S]& Y & 44 [N \& S]\\
1  &  30$\times$10 & 28 [S]& Y & 44 [N \& S]\\
1  &  30$\times$20 & 28 [S]& Y & 44 [N \& S]\\

2  & 30$\times$20 & 28 [N] & Y & 44 [N \& S]\\ 
    & 20$\times$10 & 28 [S] &  &  \\
2  & 30$\times$20 & 28 [S] & Y & 44 [N \& S]\\ 
    & 20$\times$10 & 28 [N] &  &  \\ 
2  & 30$\times$20 & 28 [N] & Y & 55 [N \& S]\\ 
    & 20$\times$10 & 28 [S] &  &  \\
2  & 30$\times$20 & 28 [S] & Y & 55 [N \& S]\\ 
    & 20$\times$10 & 28 [N] &  &  \\    
			\hline			
		\end{tabular}
	\end{center}
\end{table*}



\begin{table}[h]
	\caption{$\tau $ correlation and p-value respectively between the corrected light curve of the target and the calibration stars for night 1. In the first column we show the x and y coordinates in pixels of the calibration stars, and their correlations and p-values of their corrected light curves at each rotational angle of the Wollaston prism. }  
	\label{corr_cal_star_ord_night1}
	\centering
	\begin{center}
		\begin{tabular}{ccccccccc}
        \hline
		\hline 
			
(x centroid, y centroid)  & 0$^\circ$& & 22.5$^\circ$& &  45.0$^\circ$& & 67.5$^\circ$\\  
\hline
  Ordinary           &  $\tau$ & p-val & $\tau$ & p-val & $\tau$ & p-val & $\tau$ & p-val \\

\hline

(1492, 442)             & 0.14 & 0.49 & 0.18 & 0.37 & -0.14 & 0.49  & 0.08  & 0.69  \\
(1234, 602)             & 0.14 & 0.49 & 0.01 & 1.0 & 0.09 & 0.69 & 0.18  & 0.37   \\
(680,   592)             & 0.21 & 0.28 & -0.04 & 0.84 & 0.45 & 0.02  & 0.22  & 0.28   \\
(1279,  455)             & 0.00 & 1.0 & -0.18 & 0.37 &-0.23 & 0.23   & -0.10  & 0.62  \\
(1696,  612)             & -0.47& 0.84   & -0.26& 0.20  & -0.20& 0.32  & -0.23& 0.24  \\
\hline
Extraordinary                &  $\tau$ & p-val & $\tau$ & p-val & $\tau$ & p-val & $\tau$ & p-val \\

\hline

(1492,  532)             & -0.02 & 0.92 & 0.28 & 0.92 & 0.16 & 0.43  & -0.21  & 0.28  \\
(1234,  692)             & -0.12 & 0.56 & -0.45 & 0.02 & -0.52 & 0.05 & -0.39  & 0.04   \\
(680,   682)             & 0.20 & 0.32 & 0.03 & 0.9 & -0.03 & 0.92  & 0.05  & 0.84   \\
(1279,  545)             & 0.00 & 1.0 & -0.09 & 0.69 &-0.04 & 0.03   & 0.07  & 0.77  \\
(1696,  704)             & 0.10& 0.62   & 0.41& 0.03 & 0.41& 0.03  & -0.07& 0.77  \\

			\hline			
		\end{tabular}
	\end{center}
\end{table}



\begin{table}[h]
	\caption{$\tau $ correlation and p-value respectively between the corrected light curve of the target and the calibration stars for night 2. In the first column, we show the x and y coordinates in pixels of the calibration stars, and their correlations and p-values of their corrected light curves at each rotational angle of the Wollaston prism. }  
	\label{corr_cal_star_ord_night2}
	\centering
	\begin{center}
		\begin{tabular}{ccccccccc}
        \hline
		\hline 
			
(x centroid, y centroid)  & 0$^\circ$& & 22.5$^\circ$& &  45.0$^\circ$& & 67.5$^\circ$\\  
\hline
 Ordinary            &  $\tau$ & p-val & $\tau$ & p-val & $\tau$ & p-val & $\tau$ & p-val \\

\hline

(1280,  454) & 0.60 & 0.01 & 0.40 & 0.03 & 0.23 & 0.22 & 0.43 & 0.02 \\ 
(1492,  442) & 0.28 & 0.14 & -0.25 & 0.19 & 0.18 & 0.35 & 0.18  & 0.35 \\
(680,   592) & -0.35 & 0.06 & -0.5 & 0.01 & -0.09 & 0.62 & -0.35 & 0.06 \\
(1279,  455) & 0.25 & 0.19 & -0.19 & 0.31 & -0.03 & 0.89 & 0.41 & 0.02 \\
(1813,  623) & 0.65 & 0.01 & 0.26 & 0.16 & -0.03 & 0.89 & 0.23 & 0.22 \\
(1696,  612) & 0.05 & 0.82 & -0.19 & 0.31 & -0.01 & 0.96 & -0.31 & 0.09 \\
\hline
Extraordinary             &  $\tau$ & p-val & $\tau$ & p-val & $\tau$ & p-val & $\tau$ & p-val \\

\hline

(1280,  546) & 0.37 & 0.05 & 0.30 & 0.11 & -0.07 & 0.75 & 0.65 & 0.01 \\
(1492,  532) & -0.02 & 0.96 & -0.08 & 0.69 & -0.25 & 0.19 & -0.31 & 0.09 \\
(680,   682) & -0.43 & 0.02 & -0.35 & 0.06 & 0.03 & 0.89 & -0.57 & 0.01 \\
(1279,  545) & -0.23 & 0.23 & -0.09 & 0.62 & -0.08 & 0.69 & -0.03 & 0.89 \\
(1813,  717) & 0.06 & 0.75 & 0.08 & 0.69 & -0.30 & 0.11 & 0.33 & 0.08 \\
(1696,  704) & 0.133 & 0.51 & -0.18 & 0.35 & 0.13 & 0.51 & -0.50 & 0.01 \\

			\hline			
		\end{tabular}
	\end{center}
\end{table}


\begin{figure*}
    \centering
    \includegraphics[width=0.8\textwidth]{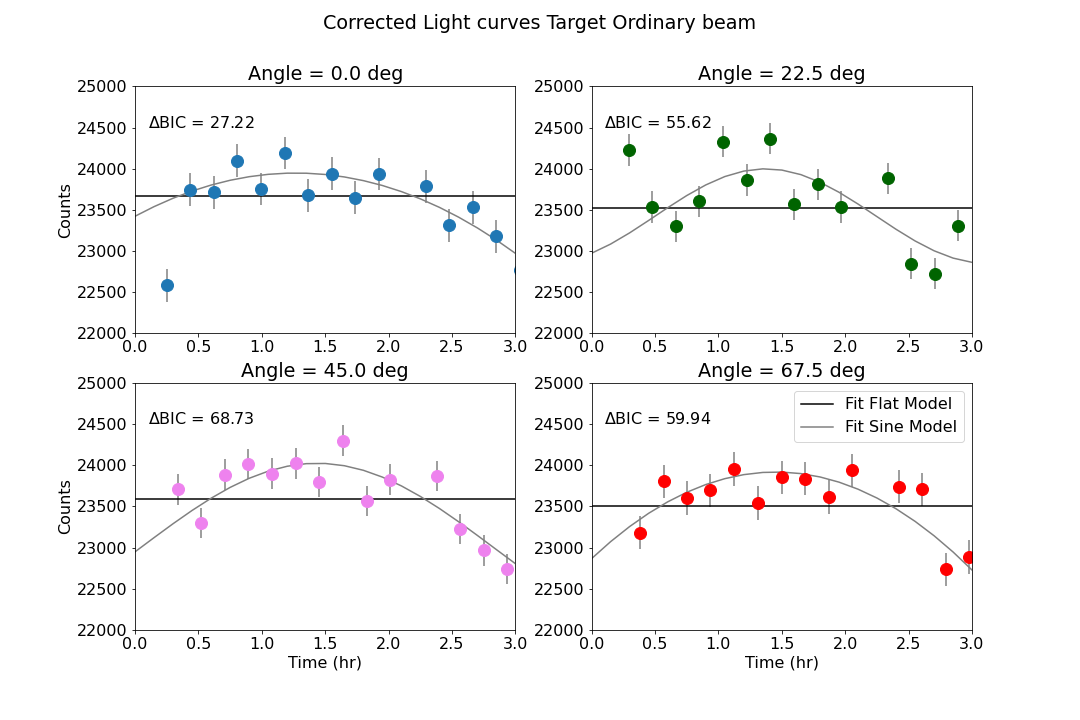}
    \includegraphics[width=0.8\textwidth]{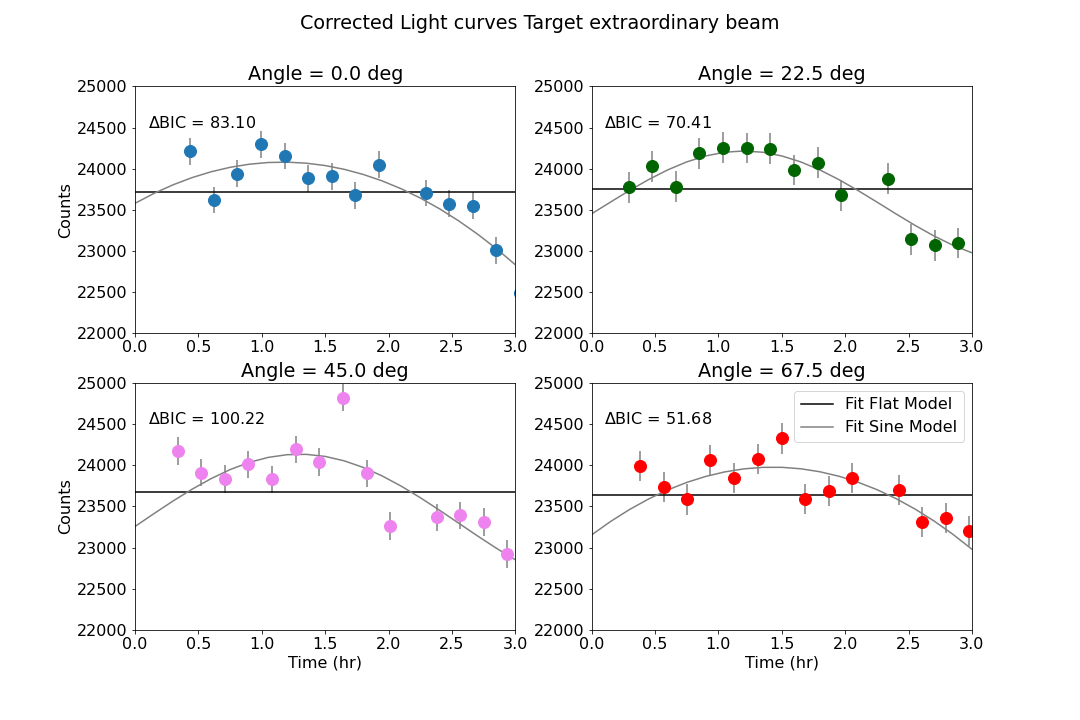}
   \caption{Corrected  light curves of 2M2139+0220 in the ordinary and extraordinary beams for angles 0.0$^\circ$, 22.5$^\circ$, 45.0$^\circ$ and 67.5$^\circ$ for night 1. {We overplot the best fitting flat model and sine model for every light curve. A positive $\Delta$BIC value indicates that the variable model is preferred over the flat model. All light curves have best fits to a sinusoidal wave.}}
    \label{ord_extra_LC_night1}
\end{figure*}

\begin{figure*}
    \centering
    \includegraphics[width=0.8\textwidth]{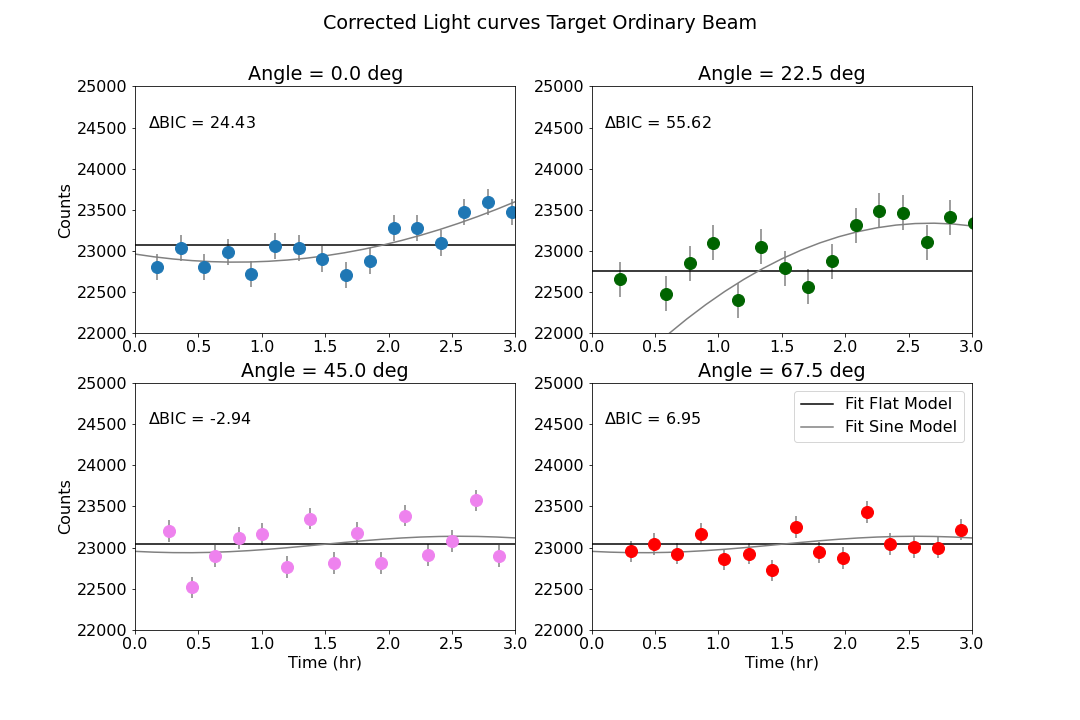}
     \includegraphics[width=0.8\textwidth]{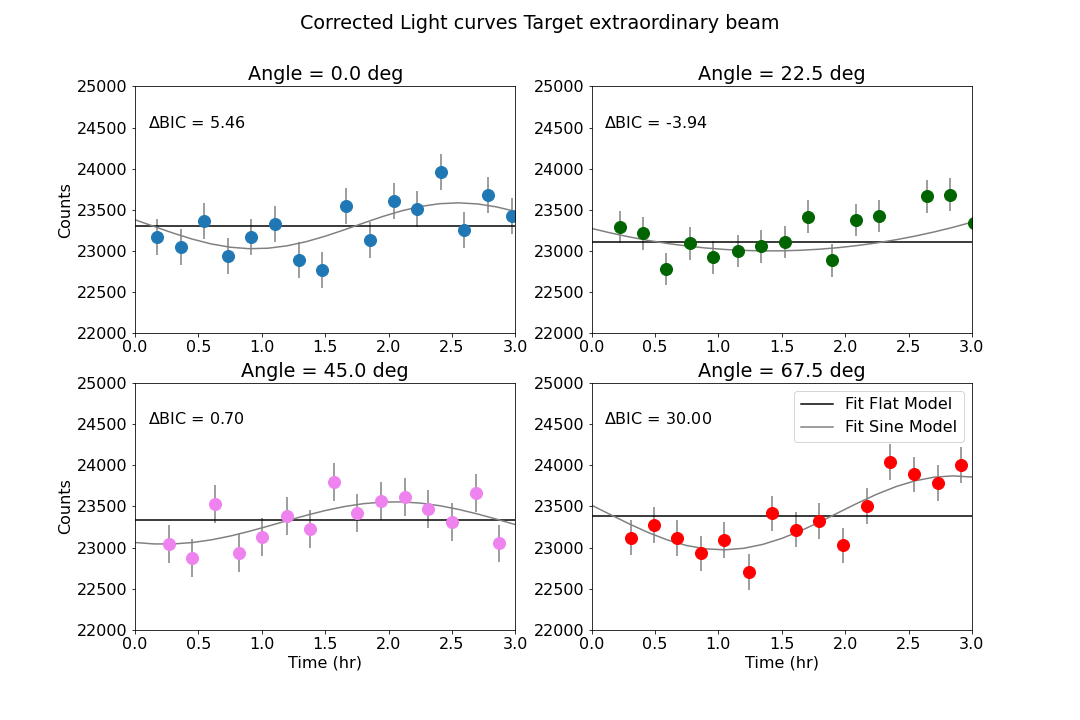}
   \caption{Corrected  light curves of 2M2139+0220 in the ordinary and extraordinary beams for angles 0.0$^\circ$, 22.5$^\circ$, 45.0$^\circ$ and 67.5$^\circ$ for night 2. {We overplot the best fitting flat model and sine model for every light curve. The ordinary light curve at 45.0 deg, and the extraordinary light curve taken at the 22.5 deg angle are best fitted by a flat line.}}
    \label{ord_extra_LC_night2}
\end{figure*}

\begin{figure*}
    \centering
    \includegraphics[width=0.8\textwidth]{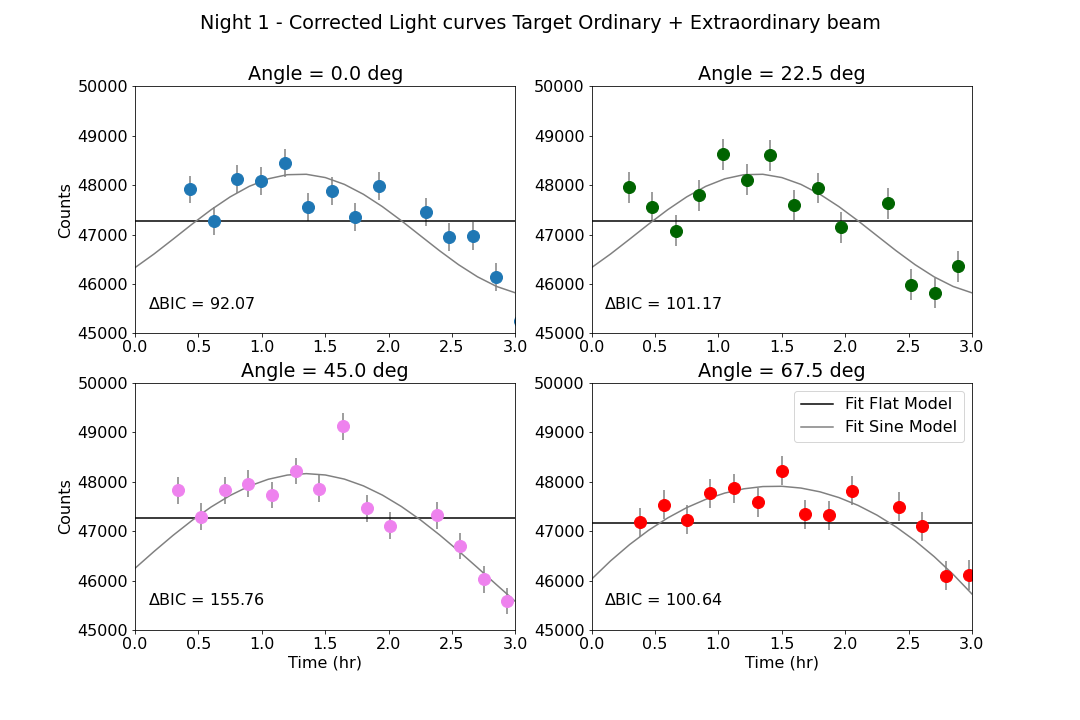}
     \includegraphics[width=0.8\textwidth]{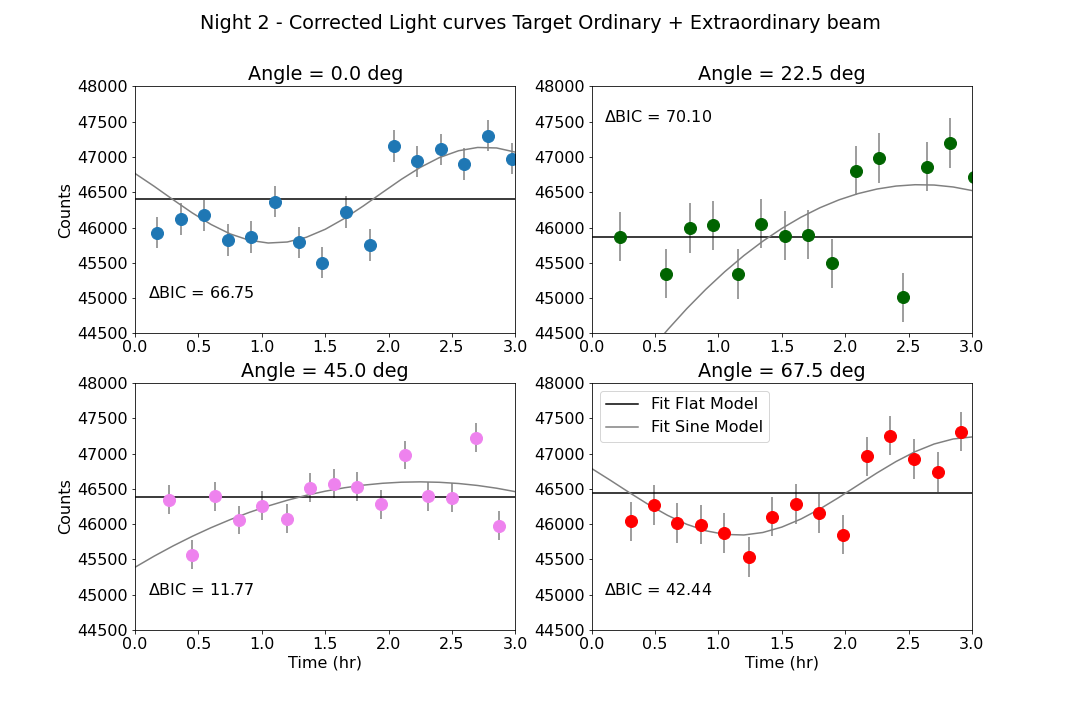}
   \caption{{Corrected  light curves of 2M2139+0220 after summing the ordinary and extraordinary beams for angles 0.0$^\circ$, 22.5$^\circ$, 45.0$^\circ$ and 67.5$^\circ$ for night 1 and 2. We overplot the best fitting flat model and sine model for every light curve. All light curves are best fitted by a sine.}}
    \label{comb_ord_extra_LC_night12}
\end{figure*}

\begin{figure*}
    \centering
    \includegraphics[width=0.95\textwidth]{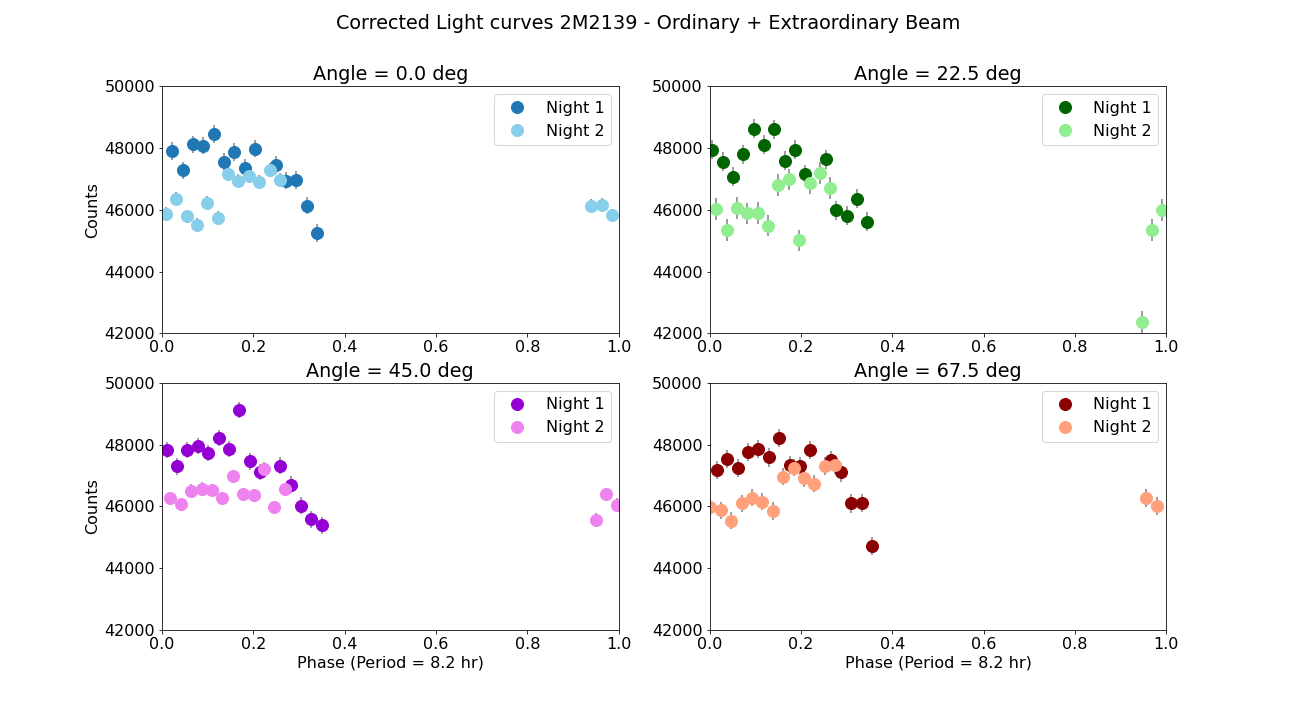}
   \caption{{Phase-foled corrected  light curves of 2M2139+0220 after summing the ordinary and extraordinary beams for angles 0.0$^\circ$, 22.5$^\circ$, 45.0$^\circ$ and 67.5$^\circ$.}}
    \label{phase_folded_comb_ord_extra_LC_night12}
\end{figure*}

\begin{figure*}
    \centering
    \includegraphics[width=0.45\textwidth]{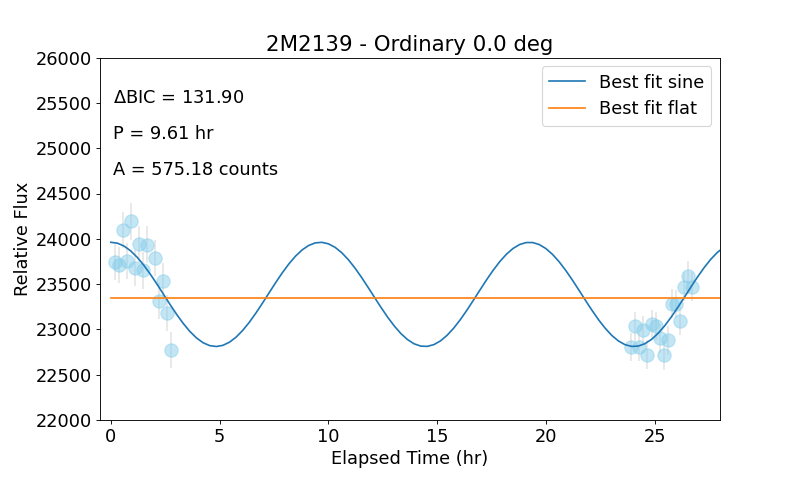}
    \includegraphics[width=0.45\textwidth]{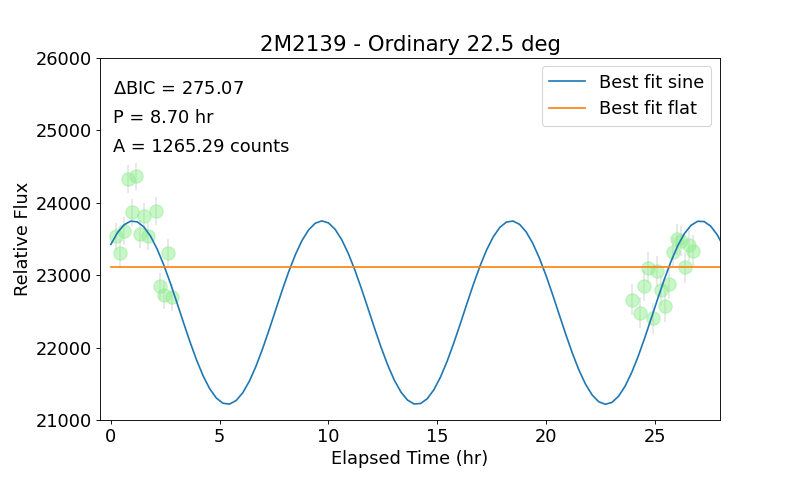}
    \includegraphics[width=0.45\textwidth]{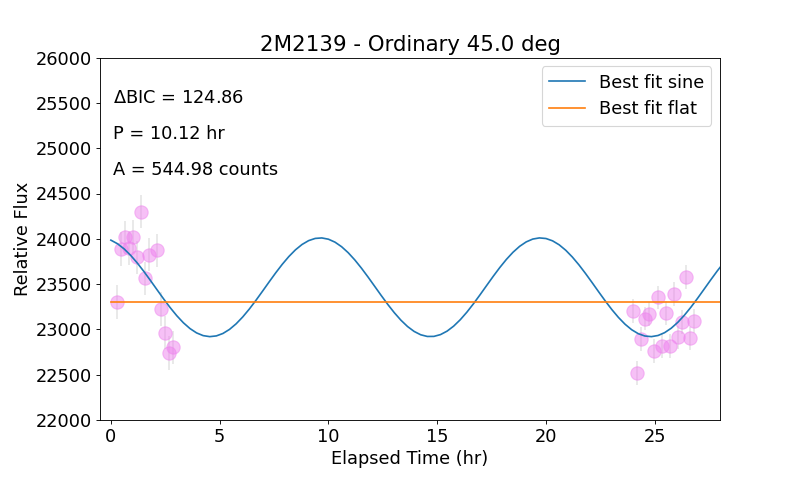}
    \includegraphics[width=0.45\textwidth]{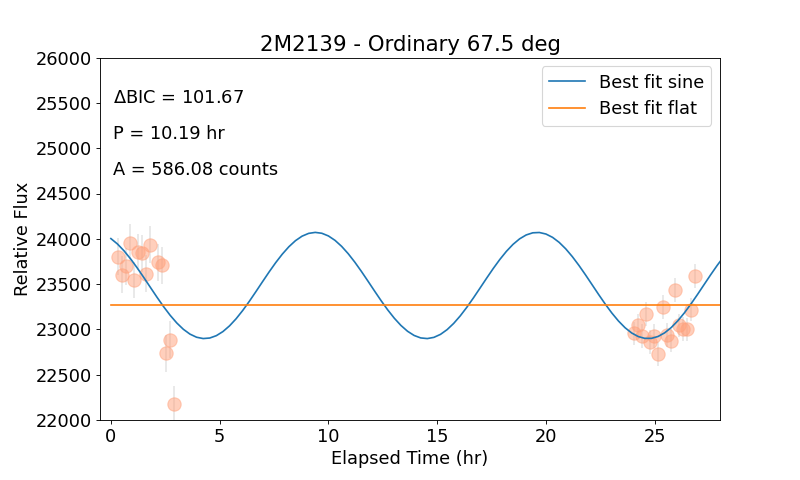}
   \caption{Corrected  light curves of 2M2139+0220 in the ordinary beams for angles 0.0$^\circ$, 22.5$^\circ$, 45.0$^\circ$ and 67.5$^\circ$ for night 1 and 2. {We overplot the best fitting flat model and sine model for every light curve. All light curves have best fits to a sinusoidal wave.}}
    \label{BIC_ord_LC}
\end{figure*}

\begin{figure*}
    \centering
    \includegraphics[width=0.45\textwidth]{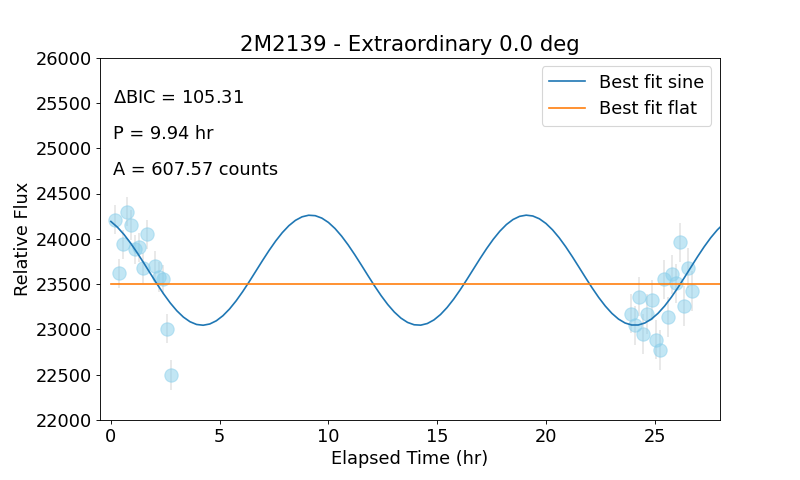}
    \includegraphics[width=0.45\textwidth]{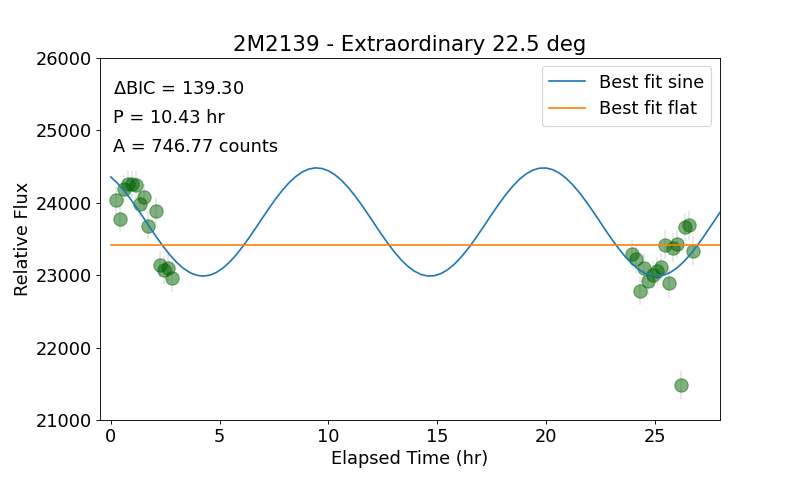}
    \includegraphics[width=0.45\textwidth]{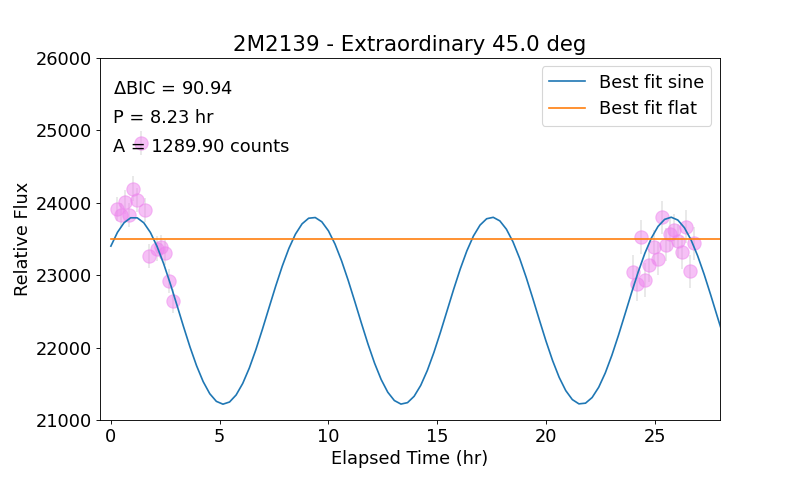}
    \includegraphics[width=0.45\textwidth]{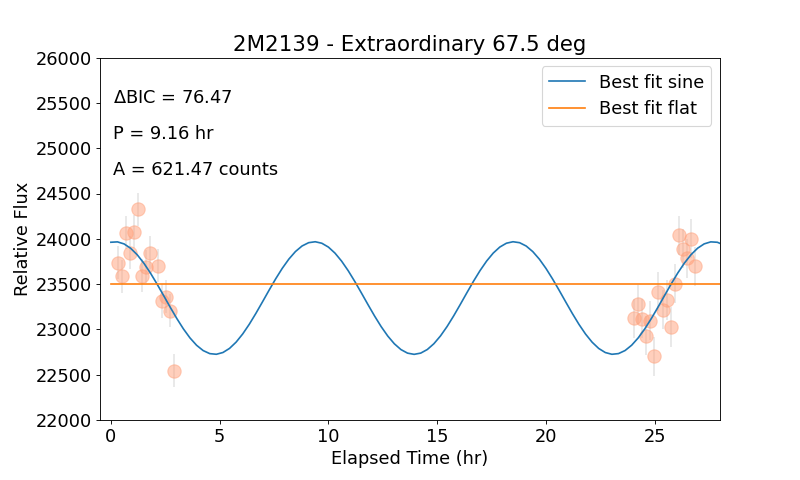}
   \caption{Corrected  light curves of 2M2139+0220 in the  extraordinary beams for angles 0.0$^\circ$, 22.5$^\circ$, 45.0$^\circ$ and 67.5$^\circ$ for night 1 and 2. {We overplot the best fitting flat model and sine model for every light curve. All light curves have best fits to a sinusoidal wave.}}
    \label{BIC_extraord_LC}
\end{figure*}

\begin{figure*}
    \centering
    \includegraphics[width=0.95\textwidth]{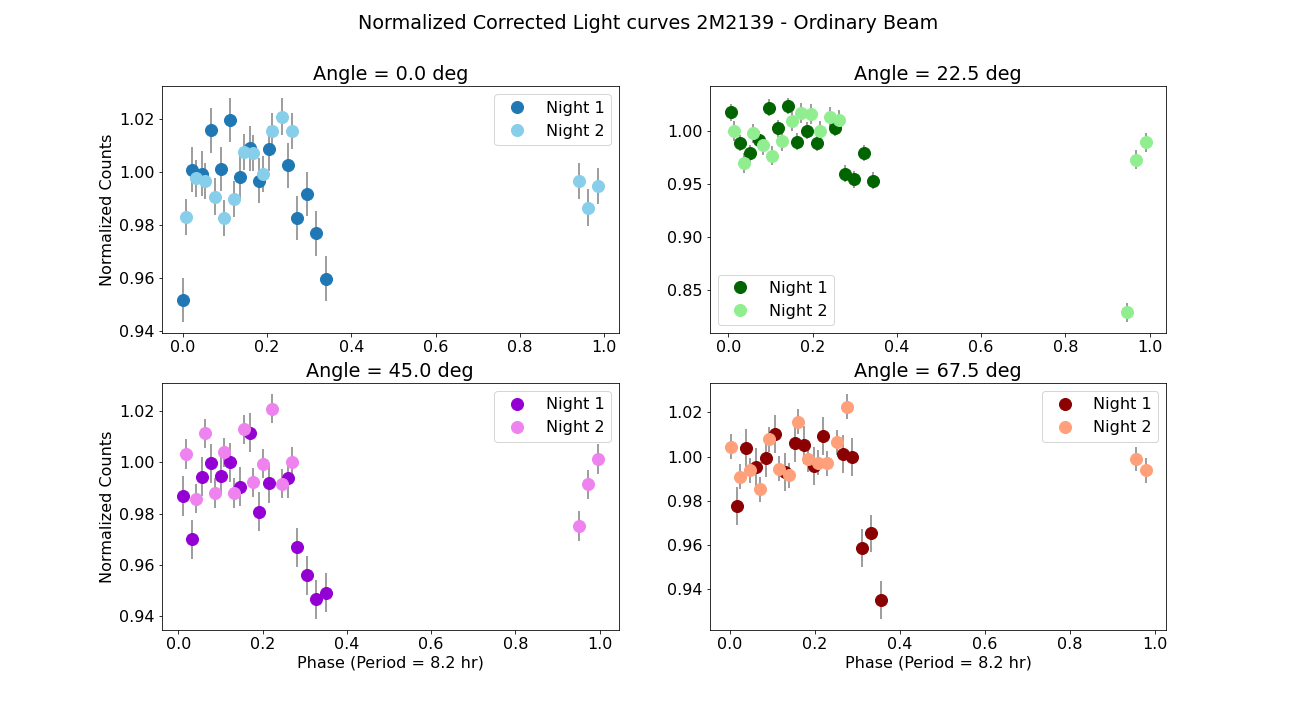}
    \includegraphics[width=0.95\textwidth]{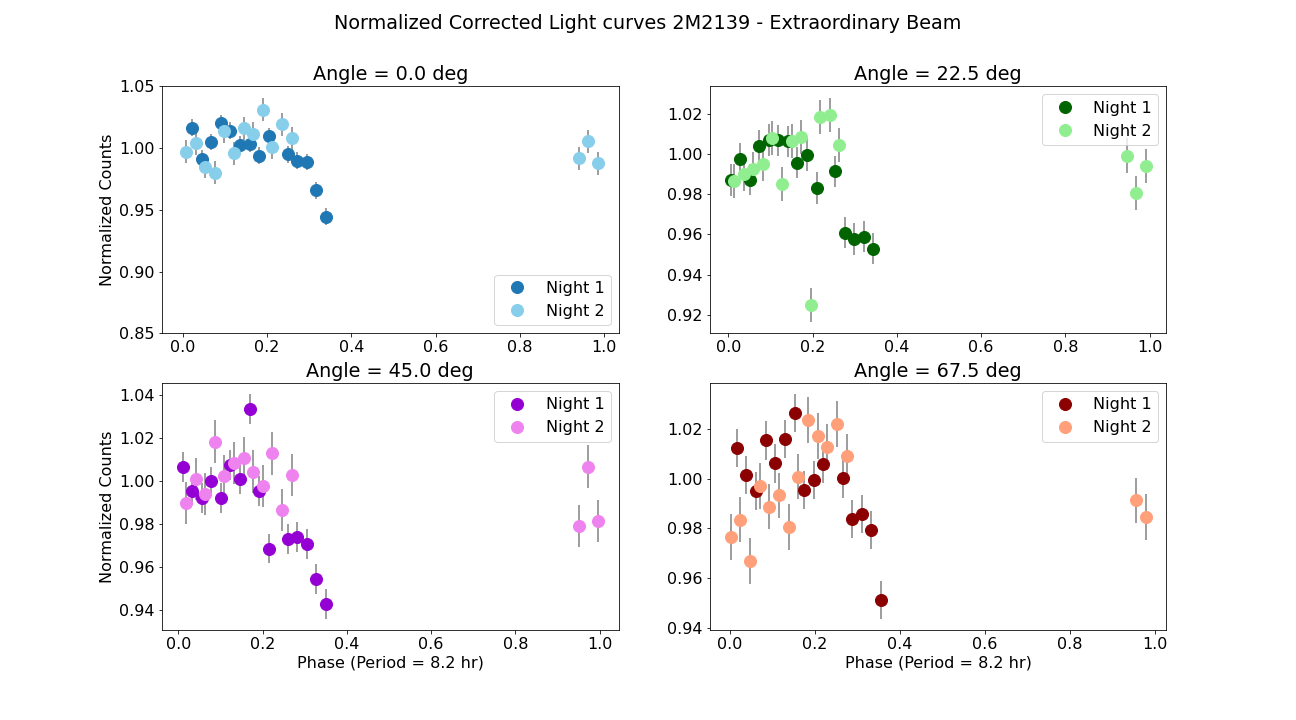}
   \caption{Normalized light curves of 2M2139+0220 in the ordinary (top) and extraordinary beams (bottom) for angles 0.0, 22.5, 45.0 and 67.5$^\circ$ for nights 1 and 2. We show the face folded light curve using the latest period from \cite{Apai2017} of 8.2~hr. }
    \label{ord_extra_LC_beams_normalized}
\end{figure*}

\begin{figure*}
    \centering
    \includegraphics[width=0.95\textwidth]{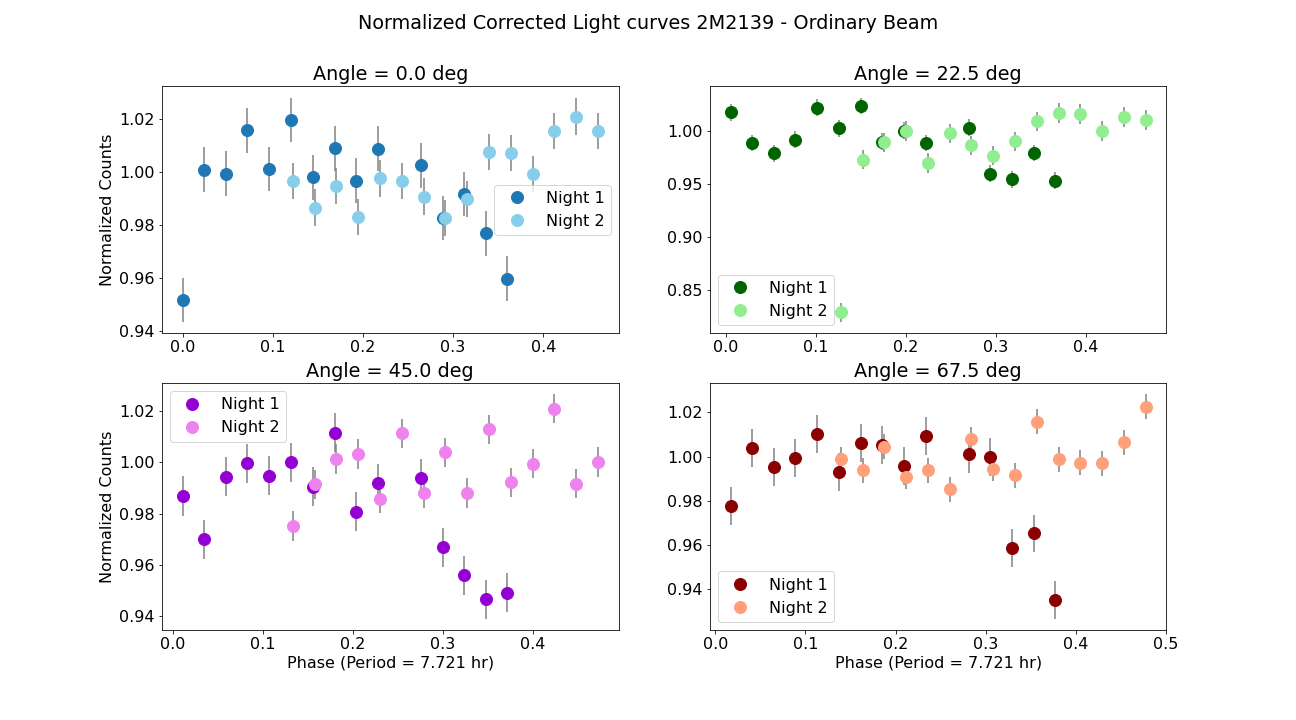}
    \includegraphics[width=0.95\textwidth]{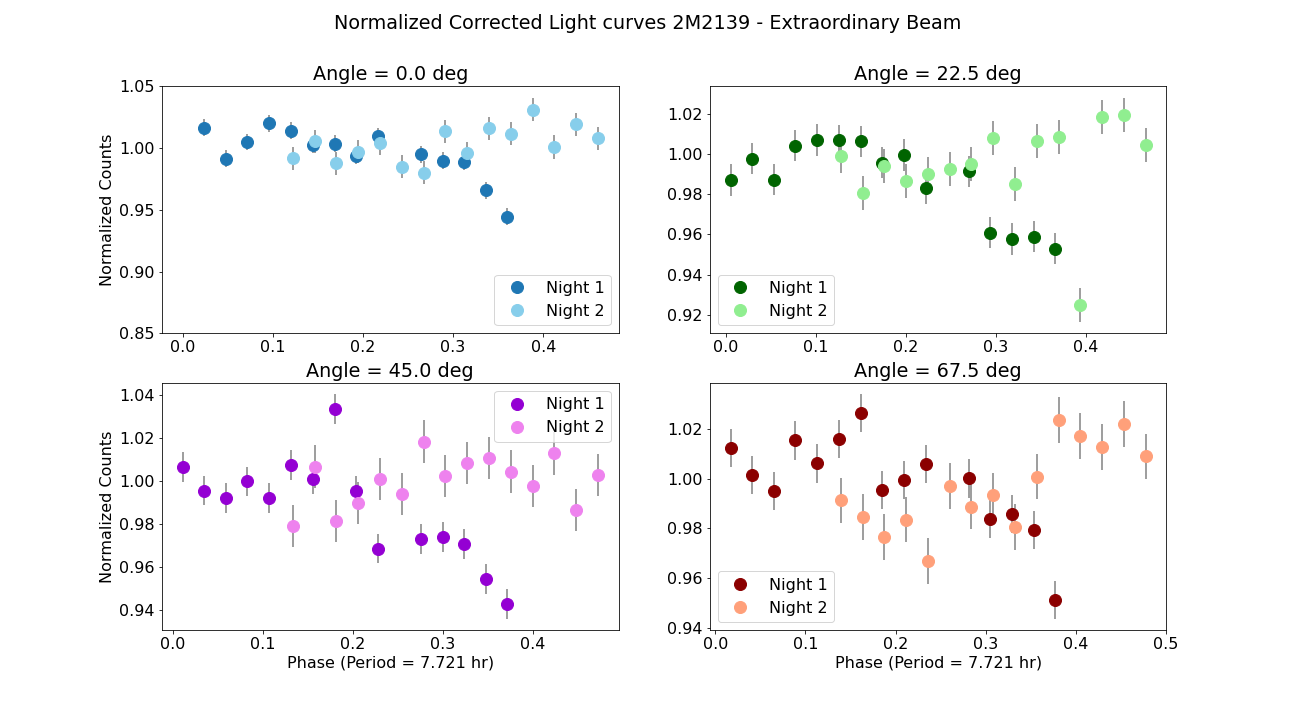}
   \caption{Normalized light curves of 2M2139+0220 in the ordinary (top) and extraordinary beams (bottom) for angles 0.0, 22.5, 45.0 and 67.5$^\circ$ for nights 1 and 2. We show the face folded light curve using the period of 7.721~hr from \cite{Radigan2012} for comparison with the previous Fig. \ref{ord_extra_LC_beams_normalized}. }
    \label{ord_extra_LC_beams_normalized_7.721}
\end{figure*}




\end{document}